\documentclass[12pt]{iopart}

\usepackage{hyperref} 
\usepackage{graphicx}
\usepackage{color}
\usepackage{dcolumn}
\usepackage{bm}
\usepackage{verbatim} 
\usepackage{cite} 
\usepackage[normalem]{ulem}

\begin{document}


\newcommand{\eric}[1]{{\color{blue}#1}}
\newcommand{\guido}[1]{{\color{violet}#1}}
\newcommand{\matthias}[1]{{\color{blue}#1}}
\newcommand{\fabian}[1]{{\color{blue}#1}}
\newcommand{\di}[1]{{\color{blue}#1}}
\newcommand{\ericC}[1]{{\color{red}\textit{\textbf{Eric:} #1}}}
\newcommand{\guidoC}[1]{{\color{red}\textit{\textbf{Guido:} #1}}}
\newcommand{\matthiasC}[1]{{\color{red}\textit{\textbf{Matthias:} #1}}}
\newcommand{\fabianC}[1]{{\color{red}\textit{\textbf{Fabian:} #1}}}
\newcommand{\diC}[1]{{\color{red}\textit{\textbf{Di:} #1}}}

\def\FileRef{
\input FName
{
\newcount\hours
\newcount\minutes
\newcount\min
\hours=\time
\divide\hours by 60
\min=\hours
\multiply\min by 60
\minutes=\time
\
\advance\minutes by -\min
{\small\rm\em\the\month/\the\day/\the\year\ \the\hours:\the\minutes
\hskip0.125in{\tt\FName}
}
}}

\mathchardef\muchg="321D
\let\na=\nabla
\let\pa=\partial

\let\muchg=\gg

\let\t=\tilde
\let\ga=\alpha
\let\gb=\beta
\let\gc=\chi
\let\gd=\delta
\let\gD=\Delta
\let\ge=\epsilon
\let\gf=\varphi
\let\gg=\gamma
\let\gh=\eta
\let\gj=\phi
\let\gF=\Phi
\let\gk=\kappa
\let\gl=\lambda
\let\gL=\Lambda
\let\gm=\mu
\let\gn=\nu
\let\gp=\pi
\let\gq=\theta
\let\gr=\rho
\let\gs=\sigma
\let\gt=\tau
\let\gw=\omega
\let\gW=\Omega
\let\gx=\xi
\let\gy=\psi
\let\gY=\Psi
\let\gz=\zeta

\let\lbq=\label
\let\rfq=\ref
\let\na=\nabla
\def\daI{{\dot{I}}}
\def\dsq{{\dot{q}}}
\def\dgj{{\dot{\phi}}}

\def\bgs{\bar{\sigma}}
\def\bgh{\bar{\eta}}
\def\bgg{\bar{\gamma}}
\def\bgy{\bar{\psi}}
\def\bgF{\bar{\Phi}}
\def\bgY{\bar{\Psi}}

\def\baF{\bar{F}}
\def\bsj{\bar{j}}
\def\baJ{\bar{J}}
\def\bsp{\bar{p}}
\def\baP{\bar{P}}
\def\bsx{\bar{x}}

\def\hgj{\hat{\phi}}
\def\hgq{\hat{\theta}}

\def\HaT{\hat{T}}
\def\HaR{\hat{R}}
\def\Hsb{\hat{b}}
\def\Hsh{\hat{h}}
\def\Hsz{\hat{z}}

\let\gG=\Gamma
\def\taA{{\tilde{A}}}
\def\taB{{\tilde{B}}}
\def\taG{{\tilde{G}}}
\def\tsp{{\tilde{p}}}
\def\tsv{{\tilde{v}}}
\def\tgF{{\tilde{\Phi}}}

\def\wgx{{\bm{\xi}}}
\def\wgz{{\bm{\zeta}}}

\def\wse{{\bf e}}
\def\wsk{{\bf k}}
\def\wsi{{\bf i}}
\def\wsj{{\bf j}}
\def\wsl{{\bf l}}
\def\wsn{{\bf n}}
\def\wsp{{\bf p}}
\def\wsr{{\bf r}}
\def\wsu{{\bf u}}
\def\wsv{{\bf v}}
\def\wsx{{\bf x}}

\def\vaB{\vec{B}}
\def\vse{\vec{e}}
\def\vsh{\vec{h}}
\def\vsl{\vec{l}}
\def\vsv{\vec{v}}
\def\vgn{\vec{\nu}}
\def\vgk{\vec{\kappa}}
\def\vgt{\vec{\gt}}
\def\vgx{\vec{\xi}}
\def\vgz{\vec{\zeta}}

\def\waA{{\bf A}}
\def\waB{{\bf B}}
\def\waD{{\bf D}}
\def\waE{{\bf E}}
\def\waF{{\bf F}}
\def\waJ{{\bf J}}
\def\waV{{\bf V}}
\def\waX{{\bf X}}

\def\R#1#2{\frac{#1}{#2}}
\def\btbl{\begin{tabular}}
		\def\etbl{\end{tabular}}
\def\bqbl{\begin{eqnarray}}
		\def\eqbl{\end{eqnarray}}
\def\ebox#1{
	\begin{eqnarray}
		#1
	\end{eqnarray}}


\def \cred#1{{\color{red}\sout{(#1)}}}
\def \cblu#1{{\color{blue}#1}}

\title[Stability impact on island]{Stability impacts from the current and pressure profile modifications within finite sized island}
\author{Yuxiang Sun$^{1}$ \& Di Hu$^{1,*}$}
\address{
	$^1$Beihang University, No. 37 Xueyuan Road, Haidian District, 100191 Beijing, China.
}
\address{
	$^*$Corresponding author.
}
\ead{hudi2@buaa.edu.cn}

\vspace{10pt}
\begin{indented}
	\item[]\today
\end{indented}

\begin{abstract}
    The stability (or instability) of finite sized magnetic island could play a significant role in disruption avoidance or disruption mitigation dynamics. Especially, various current and pressure profile modifications, such as the current drive and heating caused by electron cyclotron wave, or the radiative cooling and current expulsion caused by the Shattered Pellet Injection could be applied within the island to modify its stability, thus change the ensuing dynamics. In this study, we calculate the mode structure modification caused by such profile changes within the island using the perturbed equilibrium approach, thus obtain the change of stability criterion $\gD'$ and assess the corresponding quasi-linear island stability. The positive helical current perturbation is found to always stabilize the island, while the negative one is found to do the opposite, in agreement with previous results. The pressure bump or hole within the island has a more complicated stability impact. In the small island regime, its contribution is monotonic, with pressure bump tends to stabilize the island while pressure hole destabilizes it. This effect is relatively weak, though, due to the cancellation of the pressure term's odd parity contribution in the second derivatives of the mode structure. In the large island regime, such cancellation is broken due to the island asymmetry, and the pressure contribution to stability is manifested, which is non-monotonic. The stability analysis in this paper helps to more accurately clarify the expected island response in the presence of profile modifications caused by disruption avoidance or mitigation systems.
\end{abstract}

%
\noindent{\textbf{Keywords}: Tokamak; Magnetic Island; Tearing Mode; MHD Instability
\\[1pc]
}
%
%
\maketitle
%
%

\section{Introduction}
\label{s:Intro}
Due to the nonlinear growth of global magneto-hydrodynamic (MHD) instabilities, tokamak plasmas would eventually
experience disruptions which not only interrupt the energy output but also cause irreversible damage to the device\cite{Lehnen2015JNM}.
Such damage during the Thermal Quench (TQ) phase of disruptions is characterized by a large amount of thermal load on the Plasma Facing
Components (PFCs)\cite{Lehnen2015JNM}, while the most feared disruptive consequence in the Current Quench (CQ) phase is
the localized deposition of a large amount of runaway electrons\cite{MHDinJET}.
To avoid such damages and safe guard the continued operation of future high performance devices, various Disruption Avoidance Systems (DAS) are designed, such as the Electron Cyclotron Current Drive (ECCD) or Heating (ECRH) \cite{Zohm_1999,Giruzzi_2001,Tang2020NF,Zhang2021NF}, to suppress the instability of the dominant magnetic islands by modifying the current and pressure profile within. Facing this, Disruption Mitigation Systems (DMS) such as the Shattered Pellet Injection (SPI)\cite{Shiraki2018NF,Raman_2020,Sweeney2020NF, Xu2018FST,Li2018RSI,Baylor2019NF,Park2020FED,Jachmich_2022,Sheikh_2021,SooHwan_2021,Juhyeok_2022} or other Massive
Material Injection (MMI)\cite{Lehnen2011NF,Reux2015JNM,Pautasso2011NF,
	Hollmann2011JNM,Duan2017NF,Chen2018NF,Ding2018PST,Hollmann2019PRL} techniques are used to deplete the majority of
thermal energy via radiation loss during the TQ, and raise the density to help with runaway electron suppression during the CQ phase.

There have been a lot of recent numerical explorations\cite{Nardon_2016,Raman_2020,Ferraro_2019,Di2018NF,Kim2019POP,Hoelzl2020POP,Nardon2020NF,Izzo2020NF,Zafar2021PST,Di2021NF,PingZhu_2021}
on how the DMS would interplay with global MHD modes. Such interplay could strongly
impact the mitigation efficiency and these studies provided valuable insight into the disruption mitigation dynamics.
However, current simulations often investigate the scenario of injection into initially axisymmetric plasmas, and all MHD response
is provoked by the injection itself. In reality, it is most likely that by the time the DMS is triggered, there is already
unstable long wavelength global MHD modes, such as tearing modes, existing in the target plasma. The interplay
between such finite amplitude mode and the injected materials is of great interest, since it may have a significant impact on the
density transport and the outgoing heat flux during the mitigation process.
Specifically, the islands could be further destabilized by the helical profile modification within the O-point
\cite{Di2018NF,Di2021NF}, prompting premature triggering of the TQ before the fragments penetrate into the plasma core. Such
a scenario is detrimental to the disruption mitigation efficiency.

While expensive 3D nonlinear MHD codes could be arranged to explore the aforementioned stability impact from profile modifications caused by either the DAS and the DMS, quasi-linear stability analysis using helical
Grad-Shafranov equation \cite{hu_zakharov_2015} provides a more agile and cost-effective tool. 
In our simulation environment, which was run on a virtual machine on AMD R9 5900HX cpu, the typical convergence time ranges from $50ms$ to $300ms$, depending on the strength of profile modification. With better hardware this could be further improved. 
Though it may still not be quick enough for real-time feedback control, it should be possible to obtain a surrogate model for quick equilibria solution \cite{Lao2022,Liu2022}, which could potentially be used to satisfy the demand of the real-time control. This requires dedicated work and is part of our future work.

The justification for using such a perturbed equilibrium approach for
stability analysis is that the growth time of resistive modes is usually on the order of millisecond timescale, much longer
than the Alfv\'{e}nic timescale which is the perpendicular inertia time for magnetized plasma, so that the plasma can still be
considered as a quasi-equilibrium. The advantage of this approach is its capability of avoiding the singularity in solutions close
to the resonant surface, and accurately describing the impact of profile modification within the island on the stability
criterion $\gD'$ \cite{hu_zakharov_2015}. 

Considering a dominant finite sized island in a cylindrical analogue of the tokamak plasma with a toroidal effect multiplier $1-\R{m^2}{n^2}$ \cite{Zakharov1978NF},
the 2D temperature and pressure profile modification with helical symmetry is described and their impact on the island stability
is obtained through the calculation of $\gD'$ up to the resonant surface. Several different combinations of the current density and
pressure gradient modification are considered, and it is found that positive helical current perturbation on island O-point is
always strongly stabilizing while the negative one does the opposite, consistent with previous analytical \cite{White2015POP}  and numerical results
\cite{Di2018NF,Di2021NF}. Furthermore, the pressure gradient  modification is found to have a non-monotonic impact on the
stability depending on the island size. It should be noted that, in the scope of this paper, we only analysed simple perturbation of current density and pressure gradient without consideration of specific physical mechanism. In reality, the current and pressure profile modification could be coupled together. For example, the bootstrap current modification means certain current profile modification accompanying any given pressure profile modification. Self-consistent modelling of such profile modifications is needed for more comprehensive exploration of the island stability, and this is an immediate future development we will pursue.

The paper is arranged as the following. In Section \ref{s:Qusitheory}, we introduce our quasi-linear perturbed equilibrium
equations for the mode structure calculation considering the current and pressure gradient profile modifications. This will form
the governing equation of our numerical solution. Then in Section \ref{s:Simulation}, we describe our numerical method, set up our
simulation and demonstrate the stability impact of the aforementioned profile modifications. Finally, conclusions and discussion
are given in Section \ref{s:Conclusion}.

\section{Quasi-linear perturbed equilibrium and profile modification for finite island}
\label{s:Qusitheory}
The fundamental approach of the perturbed equilibrium method is already detailed in Ref.\,\cite{hu_zakharov_2015}, but we explicitly write out the essential parts here for the completeness of the paper.
The helical Grad-Shafranov equation for the cylindrical plasma is \cite{zakharov1986equilibrium}:
\bqbl
\lbq{eq:GFequation}
\R1r\R{\pa}{\pa r}\R{r}{1+k^2r^2}\R{\pa \gY^*}{\pa r} + \R1{r^2}\R{\pa^2 \gY^*}{\pa \gq^2}
&
=
&-\left(1-\R{m^2}{n^2}\right)\bsp'(\gY^*) 
\nonumber
\\
&&
+ \R{2k\baF}{(1+k^2r^2)^2} - \R{\baF \baF'(\gY^*)}{1+k^2r^2}.
\eqbl
Where the $k = n / (mR)$, $\baF \equiv B_z + krB_\gq$. Here $m$ and $n$ are the wave number of cylindrical plasma, and
$B_z, B_\gq$ means the magnetic induction intensity in $z, \gq$ direction respectively. Here the toroidal effect multiplier $\left(1-\R{m^2}{n^2}\right)$ is added to mimic the toroidal coupling effect \cite{Zakharov1978NF}. The helical flux $\gY^*$ is the magnetic
flux through a helical membrane, and we let $\gY^* = \gY^*_0(r) + \gy $ with $\gy$ being the helical flux perturbation. We take the current density to be ${\bf \bsj} \equiv \gm_0 {\bf j}$, $\gm_0 =
4\pi \times 10^{-7} H \cdot m^{-1}$. And $\bsp = \gm_0 p(\bgY^*) $ is the plasma pressure in the helical Grad-Shafranov
equation. And the right hand side can reflect the helical current term as $j_h = \baF \baF'(\gY^*) + (1 + k^2 r^2)\bsp'(\gY^*)$.

With the long wavelength limit $k^2r^2 \ll 1$ and use the 2D version of the current and pressure profiles, we can rewrite the
Eq.\,(\rfq{eq:GFequation}) as \cite{hu_zakharov_2015}:
\bqbl
\lbq{eq:perturbedeq1}
\R1r\R{\pa}{\pa r}r\R{\pa \gY^*}{\pa r} + \R1{r^2}\R{\pa^2 \gY^*}{\pa \gq^2}
= -j^*(\gY^*) + 2k\baF - \left(1-\R{m^2}{n^2}\right)k^2(r^2 -r^2_s)P(\gY^*), \\
\lbq{eq:perturbedeq2}
j^*(\gY^*) \equiv \left[ 1 + k^2r^2_s\left(1-\R{m^2}{n^2}\right) \right]P(\gY^*) + T(\gY^*).
\eqbl
We take $P(\gY^*) \equiv \R{dp}{d\gY^*}$ as pressure gradient, and $T(\gY^*) \equiv \R12 \R{d}{d\gY^*} (\baF)^2$. Here the 2D profile provides a more accurate marginal stability boundary compared with results obtained from simple 1D perturbed profile, with the 1D profile predicting instability while the 2D island is actually marginal stable \cite{hu_zakharov_2015}.
Combining Eq.\,(\rfq{eq:perturbedeq1}) and Eq.\,(\rfq{eq:perturbedeq2}), the helical magnetic flux $\gY^*$ can be calculated for given
boundary conditions once we know the current and pressure gradient profile as a function of $\gY^*$. We can see from these
equations that the current density perturbation will directly affect the derivative of the mode structure and thus change the
instability criterion $\gD'$. But the pressure gradient perturbation has to enter the right hand side together with the $(r^2-r_s^2)$ term. 
In a symmetric island, the contribution of the pressure gradient should thus be symmetric on both sides of the resonant surface, so
that such a symmetrical island takes no effect from this term in the calculation of $\gD'$. And only with the in-out asymmetry of the finite island could this
term have an impact on the stability. 

Separating the zeroth and the principle harmonics, 
we could write down the perturbed equilibrium equations\cite{hu_zakharov_2015}: 
\bqbl
\lbq{eq:Equilibrium00}
\R{1}{r}\R{\pa}{\pa r}\left(r\R{\pa}{\pa r}\left<\gY^*\right>_0\right)
&
=
&
-\left<\left[1+k^2r_s^2\left(1-\R{m^2}{n^2}\right)\right]P+T\right>_0
\nonumber
\\
&&
-k^2\left(r^2-r_s^2\right)\left(1-\R{m^2}{n^2}\right)\left<P\right>_0
+2k B_z
,\eqbl
\bqbl
\lbq{eq:Equilibrium10}
\R{1}{r}\R{\pa}{\pa r}\left(r\R{\pa}{\pa r}\gy\right)
&
=
&
-\left<\left[1+k^2r_s^2\left(1-\R{m^2}{n^2}\right)\right]P+T\right>_1
\nonumber
\\
&&
-k^2\left(r^2-r_s^2\right)\left(1-\R{m^2}{n^2}\right)\left<P\right>_1
+\R{m^2}{r^2}\gy
.\eqbl
Here, $\left<f\right>_0$ represents the
zeroth harmonic component of function $f$, while $\left<f\right>_1$
represents the principle harmonic $m,n$. We are only considering a single helicity $m/n$ here, since our equilibrium approach requires a given symmetric direction which is determined by the helicity. Although the multi-helicity problem is important in the triggering of the TQ, the pre-TQ phase often see a single helicity dominant mode which nonlinearly drives the other helicity modes before the TQ onset\cite{Yu_2000}. If we wish to retain the nonlinear coupling by these subdominant multi-helicity modes in the future, we should retain the principle harmonic of the nonlinear term involving the other helicity modes in the right hand side of Eq.\,(\rfq{eq:Equilibrium10}), this is a potential future development of this work. The above equations could be re-written as
\bqbl
\lbq{eq:Equilibrium01}
\R{1}{r}\R{\pa}{\pa r}\left(r\R{\pa}{\pa r}\left<\gY^*\right>_0\right)
=
-\left<j^*\right>_0
-\left<\gd j^*\right>_0
+2k B_z
,\eqbl
\bqbl
\lbq{eq:Equilibrium11}
\R{1}{r}\R{\pa}{\pa r}\left(r\R{\pa}{\pa r}\gy\right)
=
-\left<j^*\right>_1
-\left<\gd j^*\right>_1
+\R{m^2}{r^2}\gy
.\eqbl
Here we defined
\bqbl
j^*\left(\gY^*\right)
\equiv
\left[1+k^2r_s^2\left(1-\R{m^2}{n^2}\right)\right]P\left(\gY^*\right)
+T\left(\gY^*\right)
,\eqbl
\bqbl
\gd j^*
\equiv
k^2\left(r^2-r_s^2\right)\left(1-\R{m^2}{n^2}\right)P\left(\gY^*\right)
.\eqbl
With a suitable description of the 2D helical modification of the helical current density $j^*$ and the pressure gradient $P$, all terms on the Right Hand Side (RHS) of Eq.\,(\rfq{eq:Equilibrium01}) and Eq.\,(\rfq{eq:Equilibrium11}) are known, we can then obtain the perturbed helical flux $\gy$ and the corresponding $\gD'$ from given boundary condition.

The aforementioned profile modifications can be expressed as the following.
Since both the helical current density and the pressure gradient are still flux functions, we could first obtain the description of the helical flux near the resonant surface, then extract the corresponding profiles' zeroth and principle harmonics, which will be used in the RHS terms of Eq.\,(\rfq{eq:Equilibrium01}) and Eq.\,(\rfq{eq:Equilibrium11}) respectively.
The flux function near the resonant point can be represented by the following equation\cite{hu_zakharov_2015}.
\bqbl
\gY^*(x,\gq) = \gY^*_{0s} + \R{1}{2}\gY''_{0s}x^2 + (\gy_s + \gy'_sx )\cos m\gq,
\eqbl
where $x$ is the radial distance to the resonant surface, and the subscript `$s$' means the value on the resonant surface. The island semi-width \emph{w} and asymmetry \emph{s} are:
\bqbl
w = 2 \sqrt{-\R{\gy_s}{\gY''_{0s}}}, \qquad s = \R{\gy'_s}{\gY''_{0s}}.
\eqbl
We use $x_{Sl}$ and $x_{Sr}$ denote the left and right boundaries of the separatrix:
\bqbl
x_{Sl}(\gq) &= -x_{S0} - s\cos m\gq, \\
x_{Sr}(\gq) &= x_{S0} - s\cos m\gq, \\
x_{S0} &\equiv \vert w \cos^2{\R{m\gq}{2}} \vert.
\eqbl
Those equations show that $w$ is the island semi-width, while $s$ induces a periodic
asymmetry for the island along the poloidal direction. It's obvious that the asymmetry results in the outward shift of X-point and inward shift of O-point of the island, causing a corresponding asymmetry in the $(r^2-r_s^2)P$ terms discussed above.

With the island geometry known, we introduce the following flux-coordinate with the
dimension of length squared:
\bqbl
\lbq{eq:fluxequation}
\gc(\gY) \equiv \R{2(\gY-\gy_s)}{\gY''_0} = x^2 - x^2_{S0} + s^2 + 2xs\cos m\gq.
\eqbl
So that the current density profile can be represented as \cite{hu_zakharov_2015}:
\bqbl
\lbq{eq:currentdensity}
j(\gc)=\left\{
\begin{array}{ll}
	j_0 - j'_l(\sqrt{\gc+d^2_l} - d_l) & + \R12j''_l\gc,\quad(x<x_{Sl})        \\
	j_0                                & + \R12j''_i\gc,\quad(x_{Sl}<x<x_{Sr}) \\
	j_0 + j'_r(\sqrt{\gc+d^2_r} - d_r) & + \R12j''_r\gc,\quad(x>x_{Sr})        \\
\end{array}
\right.
\eqbl
where $w_l \equiv w + s = -x_{Sl}(0)$, $w_r \equiv w - s = x_{Sr}(0)$, and $d_l \equiv w_l + s$, $d_r \equiv w_r - s$.
Like-wise the pressure gradient could be described as the following:
\bqbl
\lbq{eq:pressuregradient}
P(\gc)=\left\{
\begin{array}{ll}
	P_0 - P'_l(\sqrt{\gc+d^2_l} - d_l) & + \R12P''_l\gc,\quad(x<x_{Sl})        \\
	P_0                                & + \R12P''_i\gc,\quad(x_{Sl}<x<x_{Sr}) \\
	P_0 + P'_r(\sqrt{\gc+d^2_r} - d_r) & + \R12P''_r\gc,\quad(x>x_{Sr})        \\
\end{array}
\right.
\eqbl
Here we name the different zones of the MHD by the following way. The area near the resonant surface $x_{S0}$ is a magnetic island
as the magnetic lines are reconnected to form a shape of island. Then the zones at $-d_l < x < -w_l$ and $w_r < x <d_r$ are called 
buffer zones for they are perturbed by the magnetic island. And the zones that are outside of the 2D perturbed area are
called unperturbed linear zones.

\begin{figure*}
	\centering
	\noindent
	\btbl{cc}
	\parbox{2.8in}{
		\includegraphics[scale=0.46]{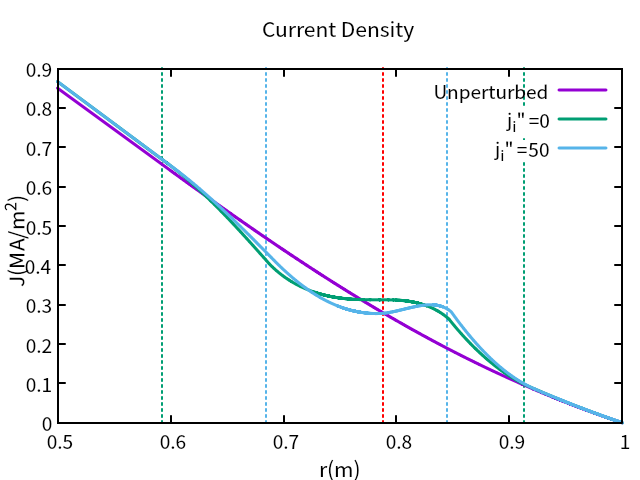}
	}
	&\quad
	\parbox{2.8in}{
		\includegraphics[scale=0.46]{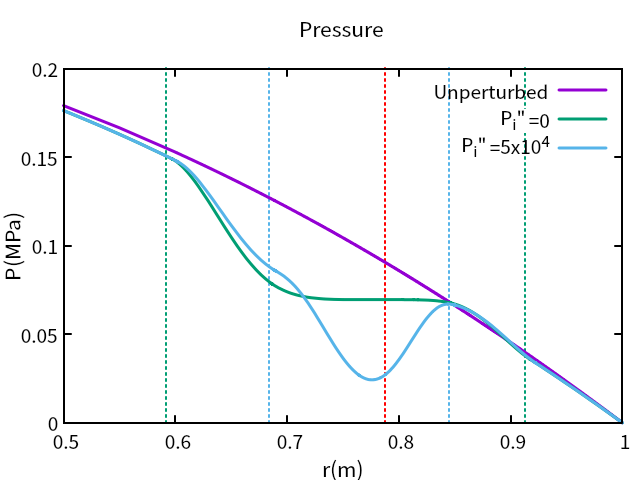}
	}
	\\
	(a)&(b)
	\etbl
	\caption{The current density and pressure of 2/1 mode with setting island semi-width to $w=0.08m$ and initial parameters to $f_1 = 0.5, f_2 = 0.35$. (a) Different $j''_i$ can change the zeroth harmony of current density greatly. The green lines are the profile of the current density which is calculated by ideal theory.
		the red dash lines and blue lines are calculated by the quasi-linear theory, while the former set $j''_2$ to zero and
		the latter one set $j''_2$ to 50. (b) As the former, shows the effect of different $P''_i$ at zeroth harmony, when the general pressure is set as 0.239 MPa. Vertical lines indicate the different zone of the island. }
	\label{fig:effect_of_ji2}
\end{figure*}

In Eq.\,(\ref{eq:currentdensity}) the parameter $ j'_l,j''_l,j'_r,j''_r $ are set to match
the value and derivative of unperturbed \textit{j} profile at the edge of the buffer zone. Besides, thanks to the fact that conductivity in the plasma is quite high, current perturbations can be limited close to the island region. So we also choose to keep the total current in quasi-linear zone the same as the unperturbed results.
And the parameter $P'_l$, $P''_l$, $P'_r$, $P''_r$ in Eq.\,(\ref{eq:pressuregradient}) are also iterated to satisfy the matching condition. Here, we consider the cooling or heating timescale to be fast compared with the global transport time, hence the pressure profiles outside of the buffer zone would not be affected significantly. A typical zeroth harmonic result of such profile modification is shown in Fig.\,\ref{fig:effect_of_ji2}, represented by the green solid lines.

In Ref \cite{hu_zakharov_2015}, the second-order gradient $j''_i$ and $P''_i$ were set to zero to represent the profile flattening within the island. As discussed in Section \ref{s:Intro}, many physical processes could result in a ``bump'' or ``hole'' on the O-point for the corresponding profiles. 
For example, the radiative cooling from impurity SPI into the
island could result in both current holes and pressure holes on the O-point \cite{Jachmich_2022,White2015POP}, while the electron cyclotron wave injection (like ECRH) could result in bumps within the island for both profiles \cite{Zohm_1999,Giruzzi_2001,Tang2020NF,Zhang2021NF}. 
More bizarre cases could exist, for example during hydrogen isotope injections,
where the dilution cooling resulted in current expulsion from the island interior but didn't deplete the pressure due to the low
radiation power. Indeed, there may even exist a minor pressure peak within the island due to the heat transfer into the low
temperature, but high density, island region.

Some examples of profile modifications with selected values of $j''_i$ and $P''_i$
are shown in Fig.\,\ref{fig:effect_of_ji2} for the zeroth harmonic which is calculated with the setting as semi-width  $w=0.08m$ and initial parameters $f_1 = 0.5, f_2 = 0.35$.
In Fig.\,\ref{fig:effect_of_ji2}(a), the unperturbed current density profile is shown in purple, a flattened profile is shown in
green and a current hole case is shown in blue. Likewise, the unperturbed, flattened and pressure hole cases are represented by
the same colours in Fig.\,\ref{fig:effect_of_ji2}(b). The red vertical dashed lines represent the location of the resonant surface, the blue dashed lines represent the maximum island width, and the green dashed lines are the 1D buffer region away from the island which is impacted by the 2D island profile modification. 

We wish to remind the readers that these profiles are merely some examples of possible profile modification caused by material injection or external drives. The exact shapes of these profile modifications are subjected to detailed analysis of heating, cooling, current drive and transport processes, which are beyond the scope of this paper. In the rest of the paper, we will perform scans on the strength of such profile modifications without delving too much into the feasibility of such modifications, as our focus is on stability analysis.

With the above description, we can finally solve the perturbed equilibrium equation from both the magnetic axis and the plasma boundary, up to both sides of the resonant surface and obtain the stability criterion
\bqbl
\gD' = \left.\R{\gy'}{\gy}\right|_{r=r_s-\ge }^{r=r_s+\ge }=\R{\gy'(r_s+\ge)-\gy'(r_s-\ge)}{\gy},\quad \ge \rightarrow 0
\eqbl
We can solve the mode structure up to the resonant surface in this approach since we avoid the singularity in the mode structure solution in the perturbed equilibrium method \cite{hu_zakharov_2015}.
We wish to emphasize here that we are calculating $\gD'$ on both sides of the reconnecting resonant surface, instead of both sides of the island boundary as conventional wisdom held \cite{Rutherford_1973,White2015POP}.
Such a $\gD'$ can be used as an indicator of the island stability. But to obtain the growth rate of the island, one may have to further carry out the flux-average of Ohm’s law carefully, which is beyond the scope of this paper.

As a further note, in toroidal geometry, one should also subtract the Pfirsh-Schluter current contribution \cite{Kotschenreuther1985} to obtain the accurate island stability. Since we are considering $\gD'$ on both side of the resonant surface, the flux-average of the Pfirsh-Schluter current on the separatrix must be used instead of the form given in Ref.\,\cite{Kotschenreuther1985}. In this study, however, due to the large major radius we chose, this additional current contribution is small \cite{Wesson2011}. Hence we simply neglect this additional contribution here, and we are satisfied with merely obtaining $\gD'$ as the indication of island stability. We will take this term into calculation in future researches to analyse the growth rate for higher inverse aspect-ration tokamak configuration.

\section{Stability impact from the current density and pressure profile modifications}
\label{s:Simulation}
\subsection{Numerical setup}
We consider a simple unperturbed equilibrium profile with the following form:
\bqbl
\lbq{eq:equilibriumprofile}
\begin{array}{rl}
	j(r) = & j(0)(1-f_2)\left[(1-f_1)\left(1-\R{r^2}{a^2}\right) + \left(1-\R{r^2}{a^2}\right)^2 \right] + \\
	       & j(0)f_2\left(1-\R{r^2}{a^2}\right)^3 ,                                                        \\
	p(r) = & p(0)\left(1-\R{r^2}{a^2}\right),
\end{array}
\eqbl
where $a$ is the minor radius. The parameter $f_1$ and $f_2$ are given manually to produce an unperturbed profile, upon which the finite island profile modification would occur. Besides those initial equilibrium profiles, there are some variables like the island semi-width $w$ and perturbation parameter $j''_i$ and $P''_i$, which we will vary in our numerical solution to find out their impact on the stability. In our numerical solution, we also choose some constant parameter values as is shown in Table \ref{tab:preset}.

\begin{table*}
	\centering
	\noindent
	\btbl{|c|c|c|}
	\hline
	Variable & Value & Physics Meaning\\
	\hline
	B (T) & 10 & Vacuum external magnetic field \\
	\hline
	a (m) & 1 & Minor radius \\
	\hline
	R (m) & 10 & Major radius \\
	\hline
	\etbl
	\caption{The initialized parameters for the quasi-linear simulation in this study. }
	\label{tab:preset}
\end{table*}

\begin{figure*}
	\centering
	\noindent
	\btbl{cc}
	\parbox{3in}{
		\includegraphics[scale=0.4]{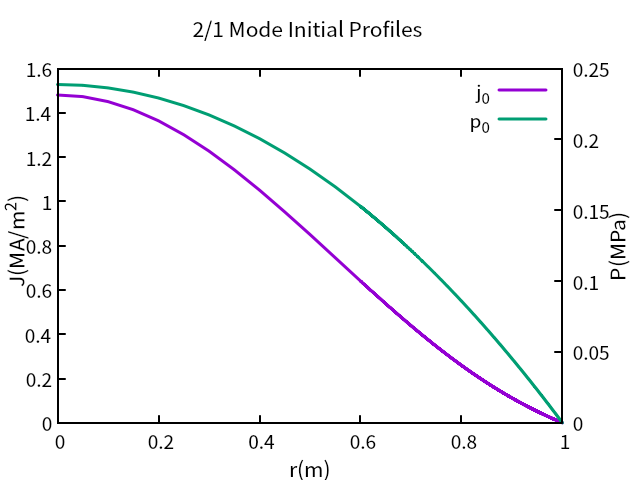}
	}
	&\quad
	\parbox{3in}{
		\includegraphics[scale=0.4]{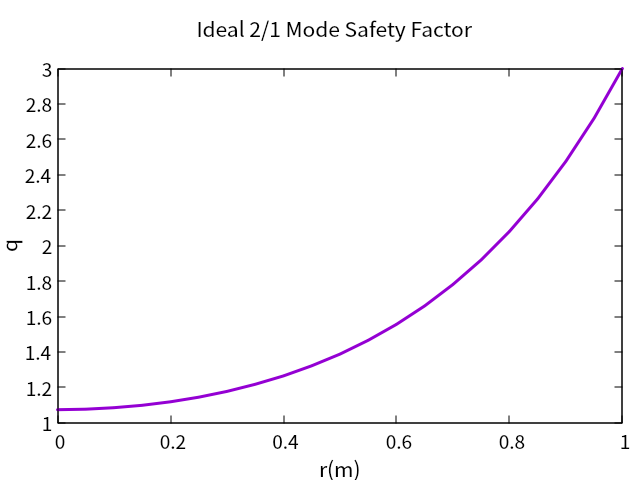}
	}
	\\
	(a)&(b)
	\\
	\parbox{3in}{
		\includegraphics[scale=0.4]{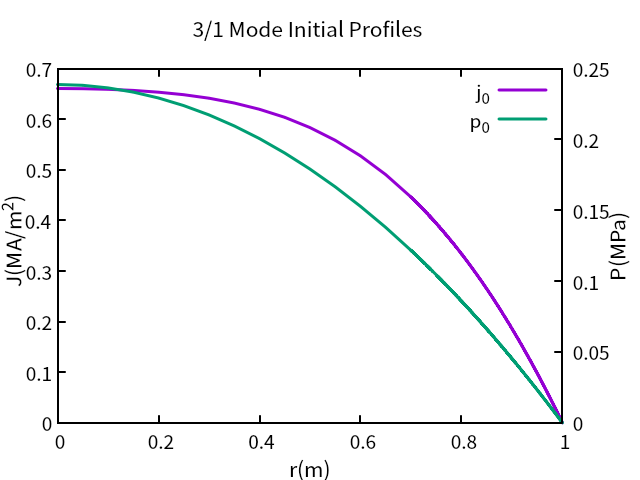}
	}
	&\quad
	\parbox{3in}{
		\includegraphics[scale=0.4]{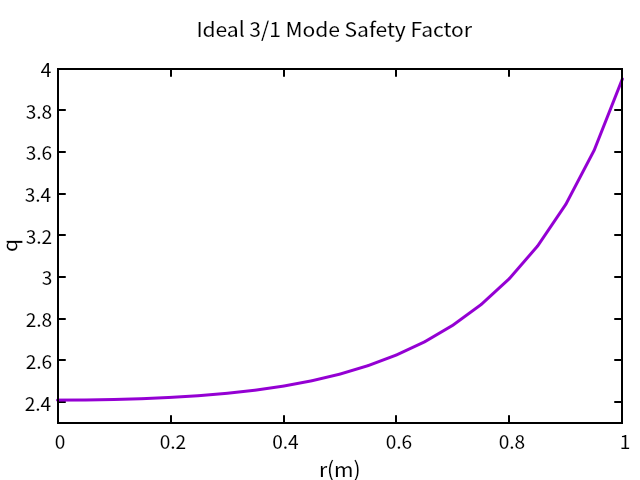}
	}
	\\
	(c)&(d)
	\etbl
	\caption{Use Eq.\ref{eq:equilibriumprofile} to calculate the initial profile used for our
		simulation, and use Eq.\ref{eq:equilibriumq} to calculate the safety factor in island.
		(a) and (b) are structures of $m = 2, n = 1$, initial parameter in Eq.\,\ref{eq:equilibriumprofile} $f_1 = 0.5, f_2 = 0.35$, edge safety factor $q(a) = 3$; (c) and (d) are structures of $m = 3, n = 1$, initial parameter in Eq.\,\ref{eq:equilibriumprofile} $f_1 = -0.30, f_2 = -0.20$,
		edge safety factor $q(a) = 3.95$.}
	\label{fig:idealprofile}
\end{figure*}

We are considering cases where there are dominant single helicity modes in the study. As an example, we investigate the $2/1$ mode and the $3/1$ mode each with a set of current profiles respectively. The $f_1$ and $f_2$ will be given for each case.
The safety factor for each current profile can be obtained by:
\bqbl
\lbq{eq:equilibriumq}
q(r) = \R{rB_{\gf}}{R_0B_{\gq}} = \R{2\pi r^2B_{\gf}}{\gm_0 I(r) R},\\
I(r) = 2\pi \int_0^rj(r')r'dr'.
\eqbl
The current density, pressure and $q$ profiles for each case are shown in Fig.\,\ref{fig:idealprofile} for the two cases respectively. The initial parameters of 2/1 mode Fig.\,\ref{fig:idealprofile}(a) is $f_1 = 0.5, f_2 = 0.35$, edge safety factor $q(a) = 3$ and the initial parameters of 3/1 mode Fig.\,\ref{fig:idealprofile}(b) is $f_1 = -0.30, f_2 = -0.20$, edge safety factor $q(a) = 3.95$.

Since Eq.\,(\rfq{eq:currentdensity}) and Eq.\,(\rfq{eq:pressuregradient}) are 2D in nature, we have to extract their zeroth and principle harmonics to be used in the RHS of Eq.\,(\rfq{eq:Equilibrium01}) and Eq.\,(\rfq{eq:Equilibrium11}). We begin with the current density profile. For regions outside of the island but within the ``buffer'' region as is marked by the green vertical dashed lines in Fig.\,\ref{fig:effect_of_ji2}, we have:
\bqbl
j^{(0)}(x) \equiv \R{1}{2\gp}\int_{-\gp}^{\gp}j(\gc)d\gq,
\eqbl
So the current density outside the island becomes:
\bqbl
\lbq{eq:currentdensity1}
j^{(0)}(x)=\left\{
\begin{array}{ll}
	j_0 - j'_lX_{1l} + \R12j''_lX^2_2 \qquad(x < x_{Sl})\\
	j_0 + j'_rX_{1r} + \R12j''_rX^2_2 \qquad(x > x_{Sr}) \\
\end{array}
\right.
\eqbl
Here we have defined:
\bqbl
X_{1l}=-d_l + \R2\gp \sqrt{d_l^2 + (x-s)^2}E\left(\R{w^2 - 4sx}{d_l^2 + (x-s)^2}\right),\\
X_{1l}=-d_r + \R2\gp \sqrt{d_r^2 + (x-s)^2}E\left(\R{w^2 - 4sx}{d_r^2 + (x-s)^2}\right),\\
X_2^2 = \R12(2x^2 + 2s^2 - w^2).
\eqbl
Here $E(x)$ is the complete elliptic integral of the second kind. 
As those equations are functions of $x$, we can calculate the zeroth harmonic of \textit{j} profile for any given radius. The perturbed current density in this buffer region is then calculated by taking the derivative of the zeroth harmonic current density just like in linear theory.

The profile modifications within the island region as is marked by the blue dashed lines in 
Fig.\,\ref{fig:effect_of_ji2} is more complicated due to its 2D nature. We have to take into account both the ``outside island region'' and the ``inside island region'' expressions in Eq.\,(\rfq{eq:currentdensity}). We define $t \equiv |x -s | / d_l$, then Eq.\,(\rfq{eq:currentdensity}) could be re-written as:
\bqbl
\lbq{eq:jprofileleft}
j_{-}(t,\gq) = j_0 - j'_ld_lF_1 + \R12d^2_lj''_lF_2 +\R12j''_iF_3w^2,\\
\lbq{eq:jprofileright}
j_{+}(t,\gq) = j_0 + j'_rd_rF_1 + \R12d^2_rj''_rF_2 +\R12j''_iF_3w^2.
\eqbl
where
\bqbl
H(t,\gq) \equiv 1,\quad\left(t>\left|\cos \left(\R{m\gq}{2}\right)\right|\right);\\
H(t,\gq) \equiv 0,\quad\left(t<\left|\cos \left(\R{m\gq}{2}\right)\right|\right);\qquad
I \equiv 1-H;\\
F_1(t,\gq) \equiv \left(\sqrt{t^2+\sin^2{\R{m\gq}{2}}-1}\right)H(t,\gq);\\
F_2(t,\gq) \equiv \left(t^2 - \cos^2{\R{m\gq}{2}}\right)H(t,\gq), \\
F_3(t,\gq) \equiv \left(t^2-\cos^2{\R{m\gq}{2}}\right)I(t,\gq)
\eqbl
It's obvious that the Eq.\,(\rfq{eq:jprofileleft}) and Eq.\,(\rfq{eq:jprofileright}) are functions of both radius $x$ and poloidal angle $\gq$ as mentioned above, to use them in Eq.\,(\rfq{eq:Equilibrium01}) and Eq.\,(\rfq{eq:Equilibrium11}), one have to extract the zeroth and principle Fourier harmonics respectively.

The pressure gradient profile can be modified like-wise. As we have discussed above, we consider the pressure increase or decrease within the island to have a faster timescale compared with that of the global transport, so that the pressure bump or hole is located within the island and its immediate vicinity. Once the pressure gradient profile $P(\gc)$ is extracted from Eq.\,(\rfq{eq:pressuregradient}), the perturbed flux could be solved. The pressure profile itself can be obtained by simply taking the integration from the boundary condition $p(a)=0$:
\bqbl
dp = P d\gY^* = P dx \cdot \R{d\gY^*}{dx}, \\
\left<p\right>_0 
= 
\int{\left<P\right>_0 \left<\R{d\gY^*}{dx} \right>_0 dx}
+\int{\left<P\right>_1^* \R{d\gy}{d x} dx}.
\eqbl
Here $\left<P\right>_1^*$ marks the conjugation of the principle harmonic of the pressure gradient. The contribution from the $\left<P\right>_1^* \R{d\gy}{d x}$ term is small compared with the average field term for our parameters considered, however. An example of the final pressure profile thus obtained has been shown in Fig.\,\ref{fig:effect_of_ji2}.

\subsection{Impact from current density profile modification within the island}

\begin{figure*}
	\centering
	\noindent
	\includegraphics[scale=0.4]{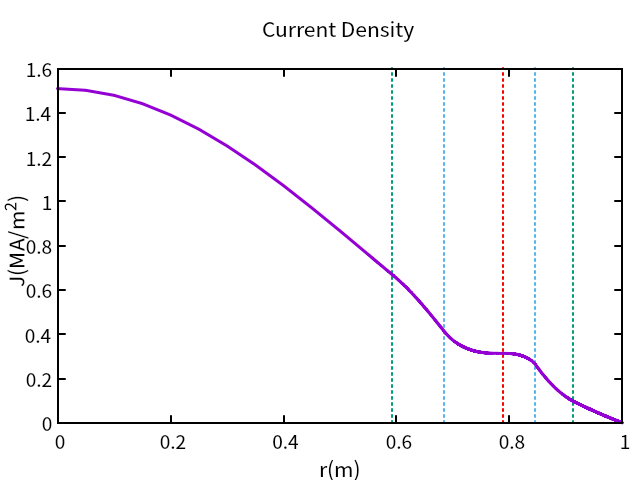}
	\\
	(a)
	\btbl{cc}
	\parbox{3in}{
		\includegraphics[scale=0.4]{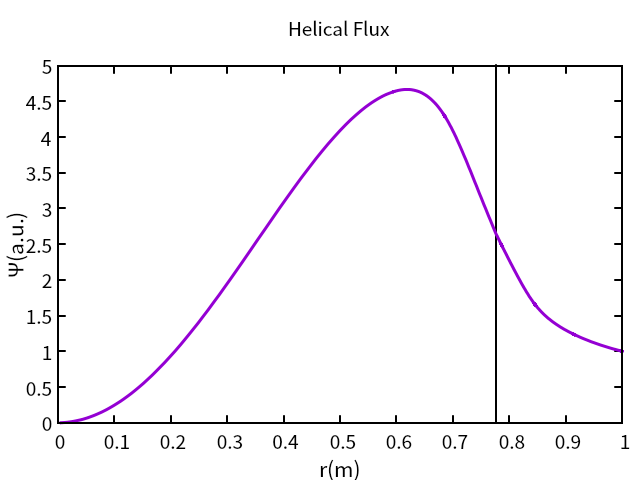}
	}
	&\quad
	\parbox{3in}{
		\includegraphics[scale=0.4]{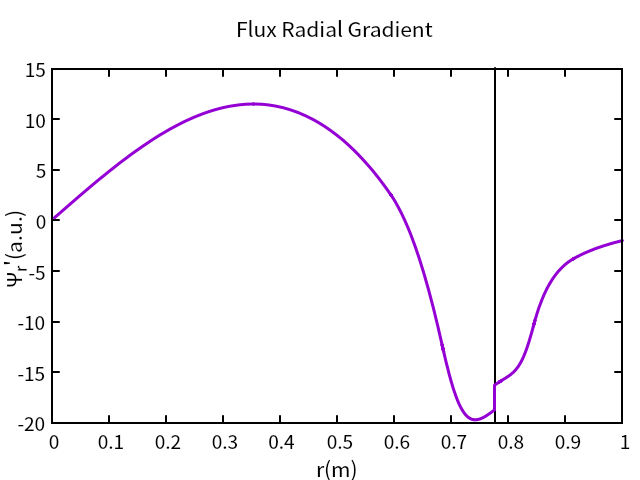}
	}
	\\
	(b)&(c)
	\etbl
	\caption{An example of the 2/1 mode structure for an island semi-wide $w=0.0801m$, initial parameter $f_1 = 0.5, f_2 = 0.35$ and without pressure ingredient. (a) The unperturbed current density
		calculated by a quasi-linear way with five vertical lines indicating the different zones of the island.  (b) the helical flux $\gy$,
		and (c) its radial gradient $\pa_r \gy$. }
	\label{fig:profileof21jnop1}
\end{figure*}

In this subsection, we first demonstrate the stability impact from only the current profile modification, and show that our perturbed equilibrium result is consistent with previous finite island analyses using constant $\gy$ assumptions on island boundary. Fig.\ref{fig:profileof21jnop1} shows an example profile we got by our code which has the island semi-width  $w=0.0801m$ and no pressure. Fig.\ref{fig:profileof21jnop1}(a)
presents the current density which is flattened within the magnetic island. As the figure indicated,
the blue dashed lines mark the edge of the magnetic island, and the green dashed lines divide the area where we need to consider the 2D perturbations, buffer zones are the areas which between the blue lines and green lines.
While the red dashed line in the middle shows the place of the resonant surface. Those vertical dashed lines have the same meaning in the following figures. What's more, Fig.\ref{fig:profileof21jnop1}(b) and (c) show the calculated magnetic flux and its first derivative. A notable feature is that the perturbed flux is not constant-$\gy$ across the island region. The jump in the $\gy_r'$ corresponds to the stability criterion $\gD'$.

\begin{figure*}
	\centering
	\noindent
	\btbl{cc}
	\parbox{3in}{
		\includegraphics[scale=0.8]{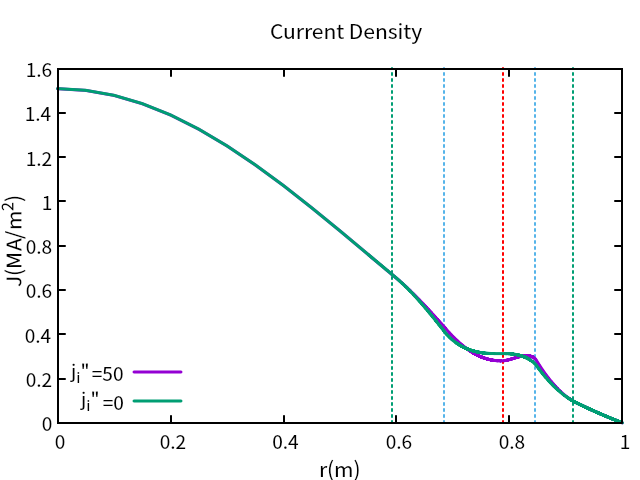}
	}
	&\quad
	\parbox{3in}{
		\includegraphics[scale=0.4]{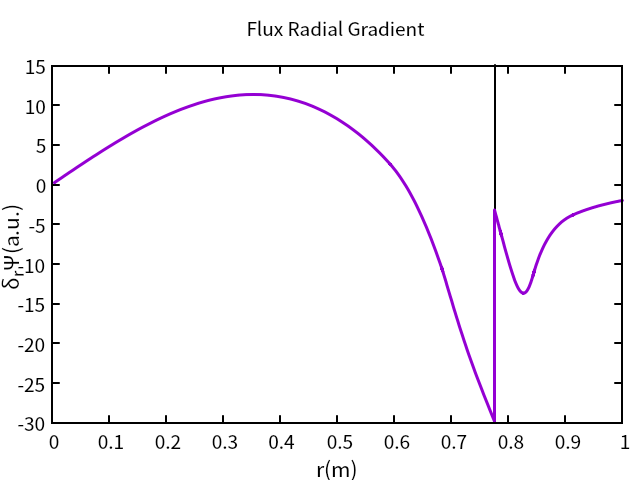}
	}
	\\
	(a)&(b)
	\\
	\parbox{3in}{
		\includegraphics[scale=0.8]{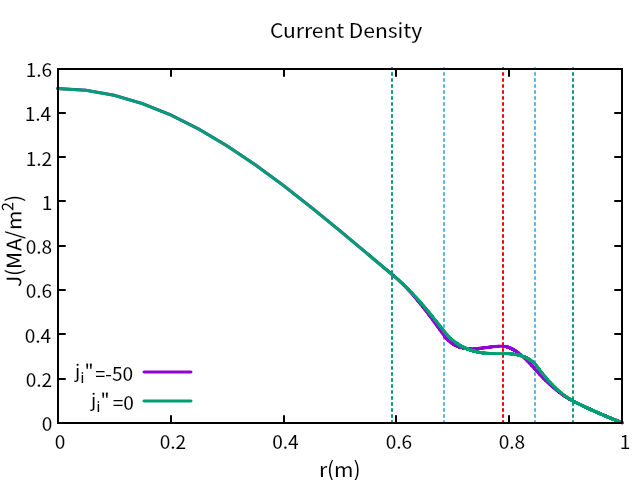}
	}
	&\quad
	\parbox{3in}{
		\includegraphics[scale=0.4]{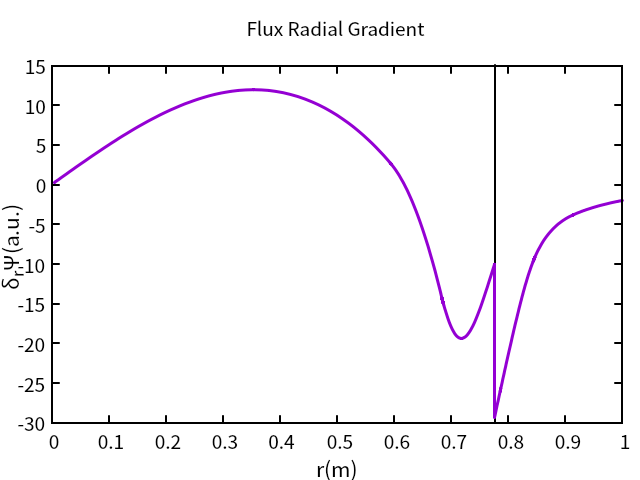}
	}
	\\
	(c)&(d)
	\etbl
	\caption{The current density of 2/1 mode with setting island semi-width to $w=0.08m$ and initial parameter $f_1 = 0.5, f_2 = 0.35$. The five vertical lines still indicate the five zones
		of the island, and the green line in each figure is the current density for perturbation is zero and the purple line is
		for the profile calculated with  (a) perturbation $j_i''$ to 50;(c) set the perturbation $j_i''$ to -50. (b) and (d) is the $\pa_r \gy$
		for the structures (a) and (c). (In figures (a) and (c), the red lines mark the resonant surface, the blue lines divide the magnetic island and the green lines are the edge of the buffer zones. In figures (b) and (d) the black lines mark the resonant surface.)}
	\label{fig:profileof21jnop2}
\end{figure*}

To demonstrate the effect of the current bump or hole, we show two further examples in Fig.\,\ref{fig:profileof21jnop2}.
As we can see, if we give a negative helical current perturbation to the island, the $\gD'$ increases, which means the instability grows. On the other hand, if we give a positive perturbation, the $\gD'$ will decrease to negative and indicates the island becomes more stable. This is consistent with previous findings in both the ECCD island stabilizing studies and the cooling island destabilizing studies\cite{Jachmich_2022,White2015POP,Zohm_1999,Giruzzi_2001}, that a positive helical current perturbation on the island O-point tends to stabilize the island, while a negative helical current perturbation on the O-point tends to do the opposite. 

\begin{figure*}
	\centering
	\noindent
	\btbl{cc}
	\parbox{3in}{
		\includegraphics[scale=0.3]{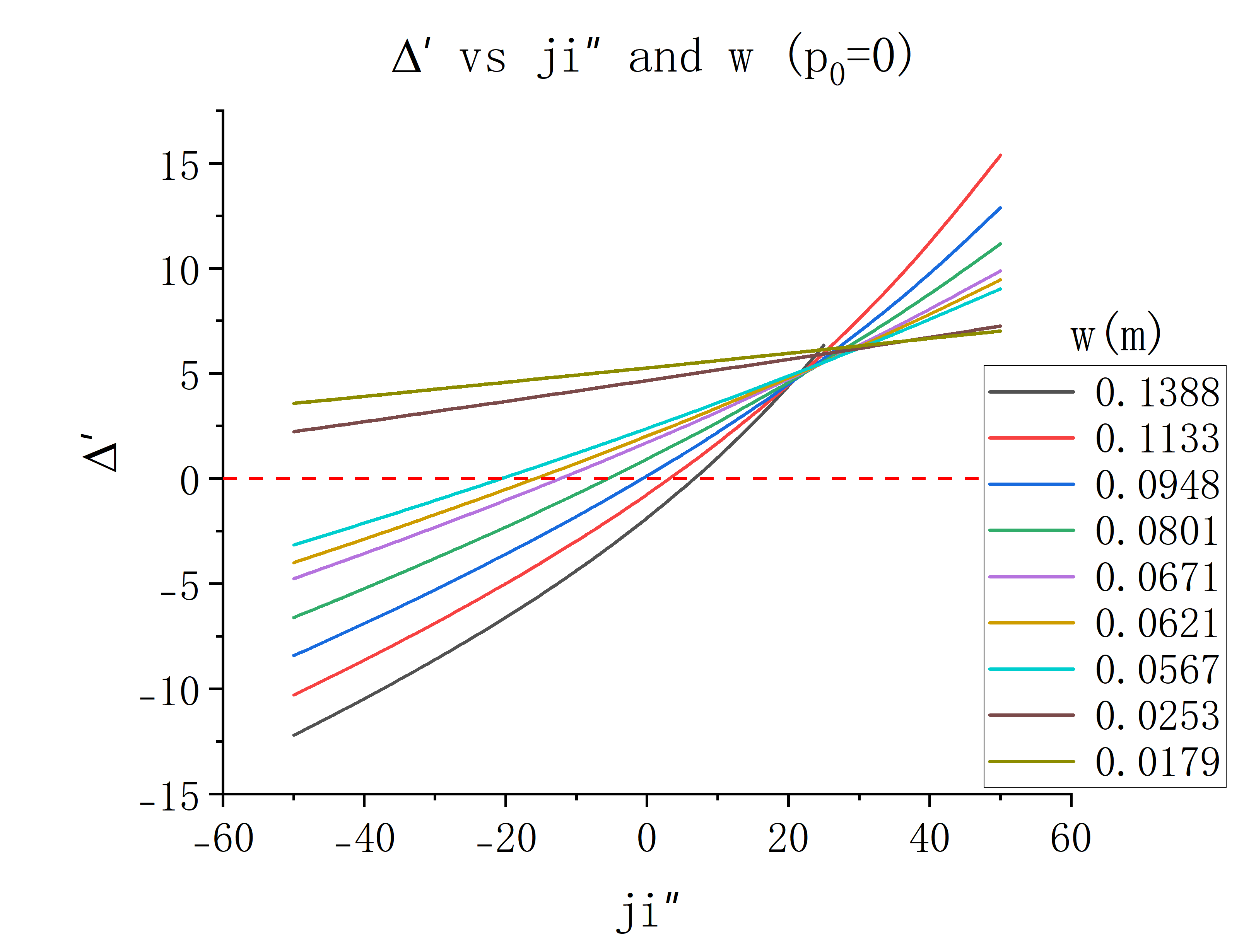}
	}
	&\quad
	\parbox{3in}{
		\includegraphics[scale=0.3]{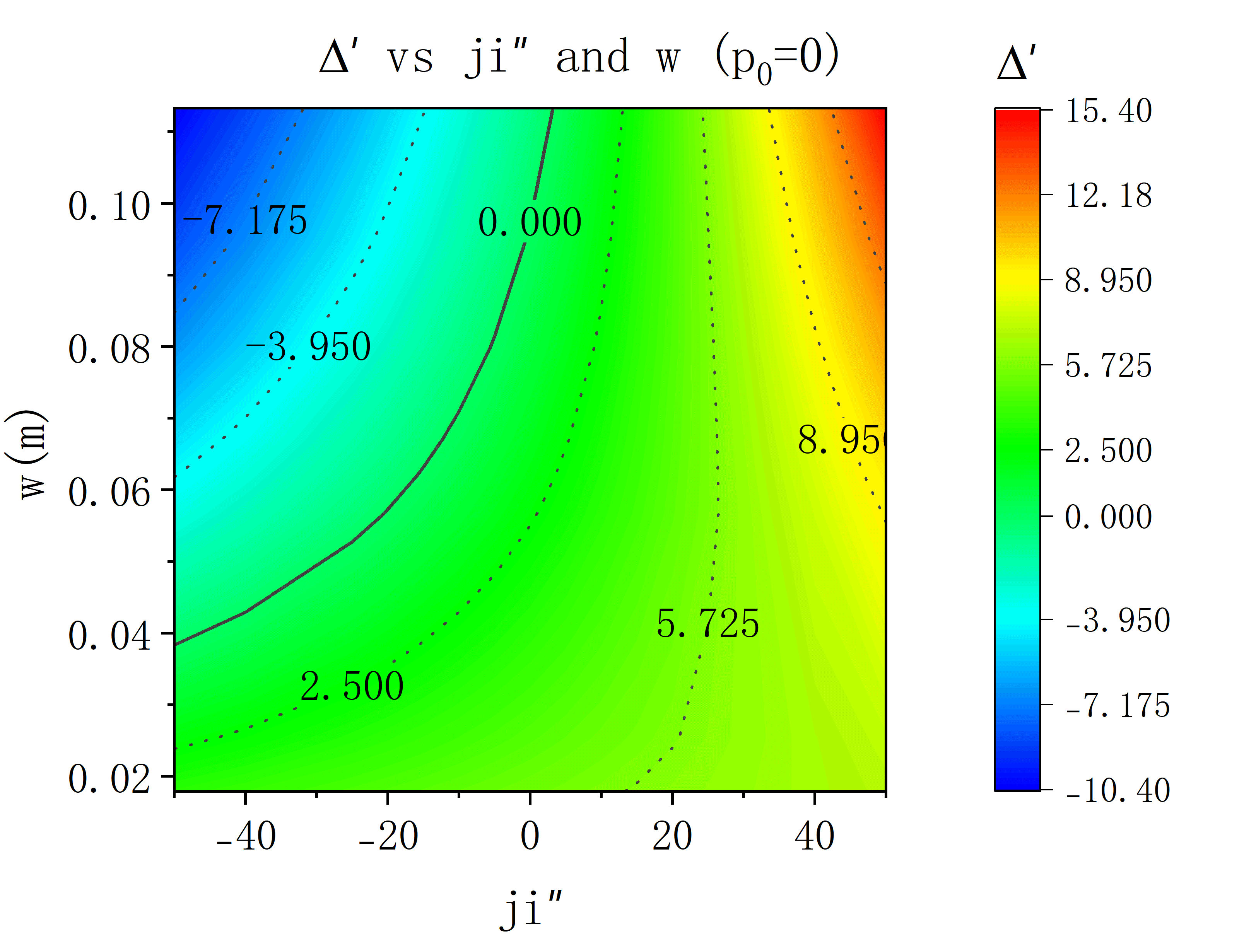}
	}
	\\
	(a)&(b)
	\\
	\parbox{3in}{
		\includegraphics[scale=0.3]{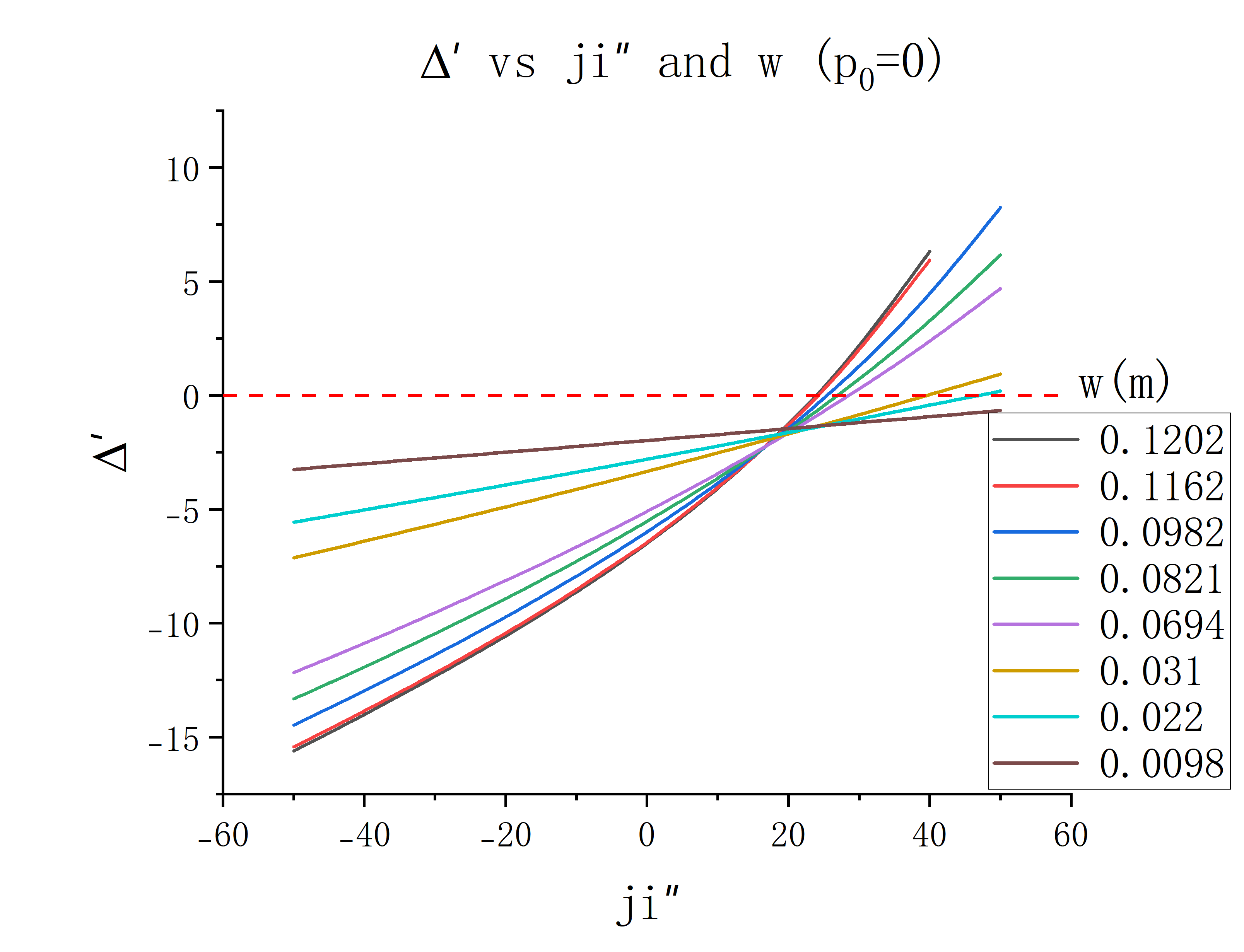}
	}
	&\quad
	\parbox{3in}{
		\includegraphics[scale=0.3]{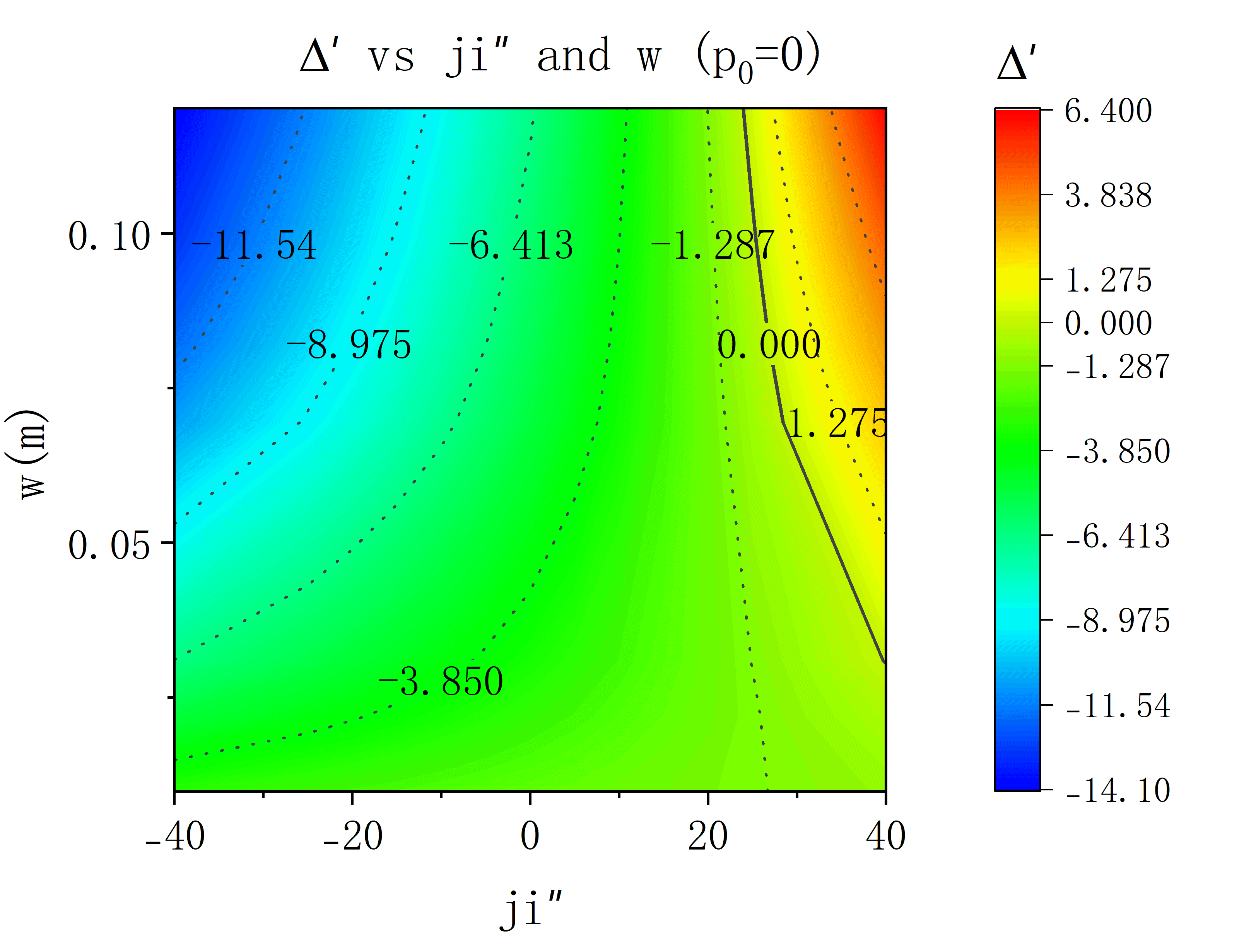}
	}
	\\
	(c)&(d)
	\etbl
	\caption{Value of $\gD' $ vary by semi-width $w$ and perturbations $j_i''$. (a) and (b) Results of 2/1 mode with edge safety factor $q(a) = 3$ and initial parameter $f_1 = 0.5, f_2 = 0.35$;
		(c) and (d) Results of 3/1 mode with edge safety factor $q(a) = 4.5$ and initial parameter $f_1 = 0.5, f_2 = 0.35$}
	\label{fig:profileofjnop}
\end{figure*}

A more comprehensive look at the helical current perturbations impact can be obtained from a parameter scan of island semi-width and current perturbation amplitude, the result of which is shown in Fig.\ref{fig:profileofjnop}, assuming no pressure contribution. In Fig.\ref{fig:profileofjnop}(a) \& (c), we show a scan of current perturbation strength with different fixed island width. 
First, as the figure indicates, the stronger the perturbation, the bigger the impact to $\gD'$. Consistent with the results in Fig.\,\ref{fig:profileof21jnop2}, the helical current bump (negative $j''_i$) results in a decrease of $\gD'$ and stabilizes the island, while the helical current hole does the opposite. Second, the effect of the current bump or hole depends on the island width, so that for the same $j''_i$, the wider the island, the stronger the effect. This is understandable as with a same $j''_i$ the more current are perturbed in the larger island, thus the stronger impact. Such a trend is common for both modes. In Fig.\ref{fig:profileofjnop}(b) \& (d), these trends are further confirmed by the contour of $\gD'$ for different island semi-width $w$ and $j''_i$. In these figures, the marginal stable boundary $\gD'=0$ are marked by a solid line, while the other contours are in dashed lines.

\begin{figure*}
	\centering
	\noindent
	\btbl{cc}
	\parbox{3in}{
		\includegraphics[scale=0.3]{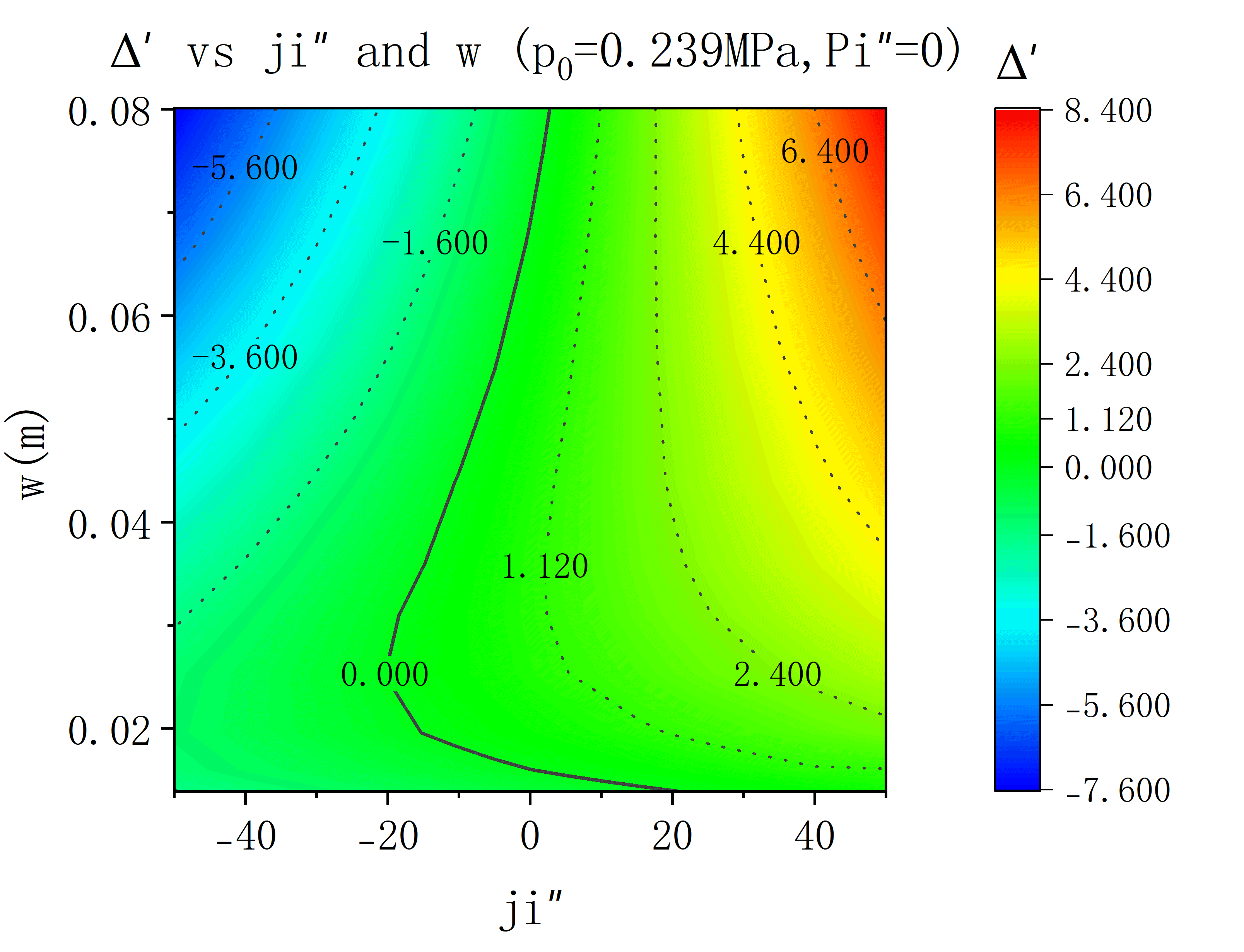}
	}
	&\quad
	\parbox{3in}{
		\includegraphics[scale=0.3]{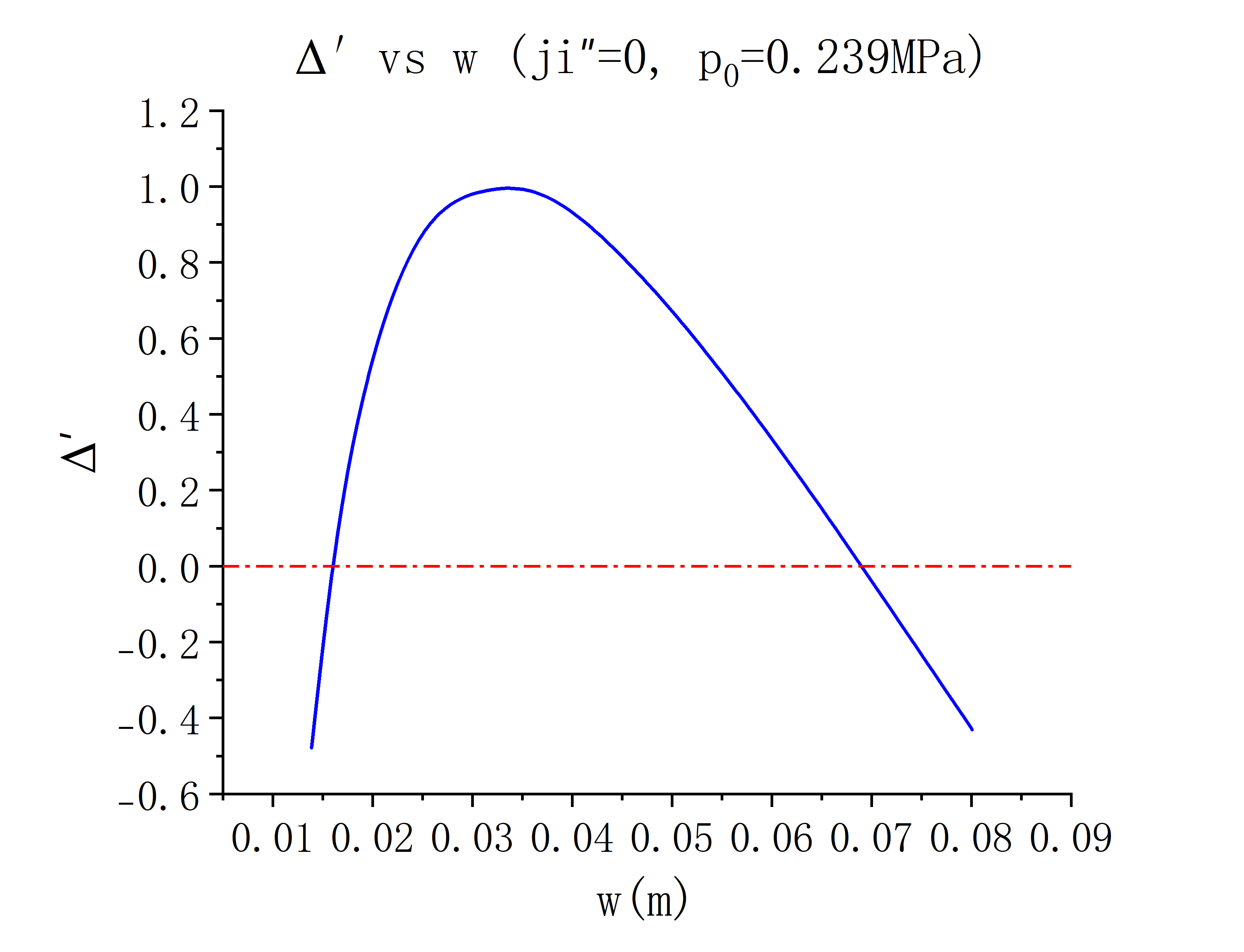}
	}
	\\
	(a)&(b)
	\etbl
	\\
	\includegraphics[scale=0.3]{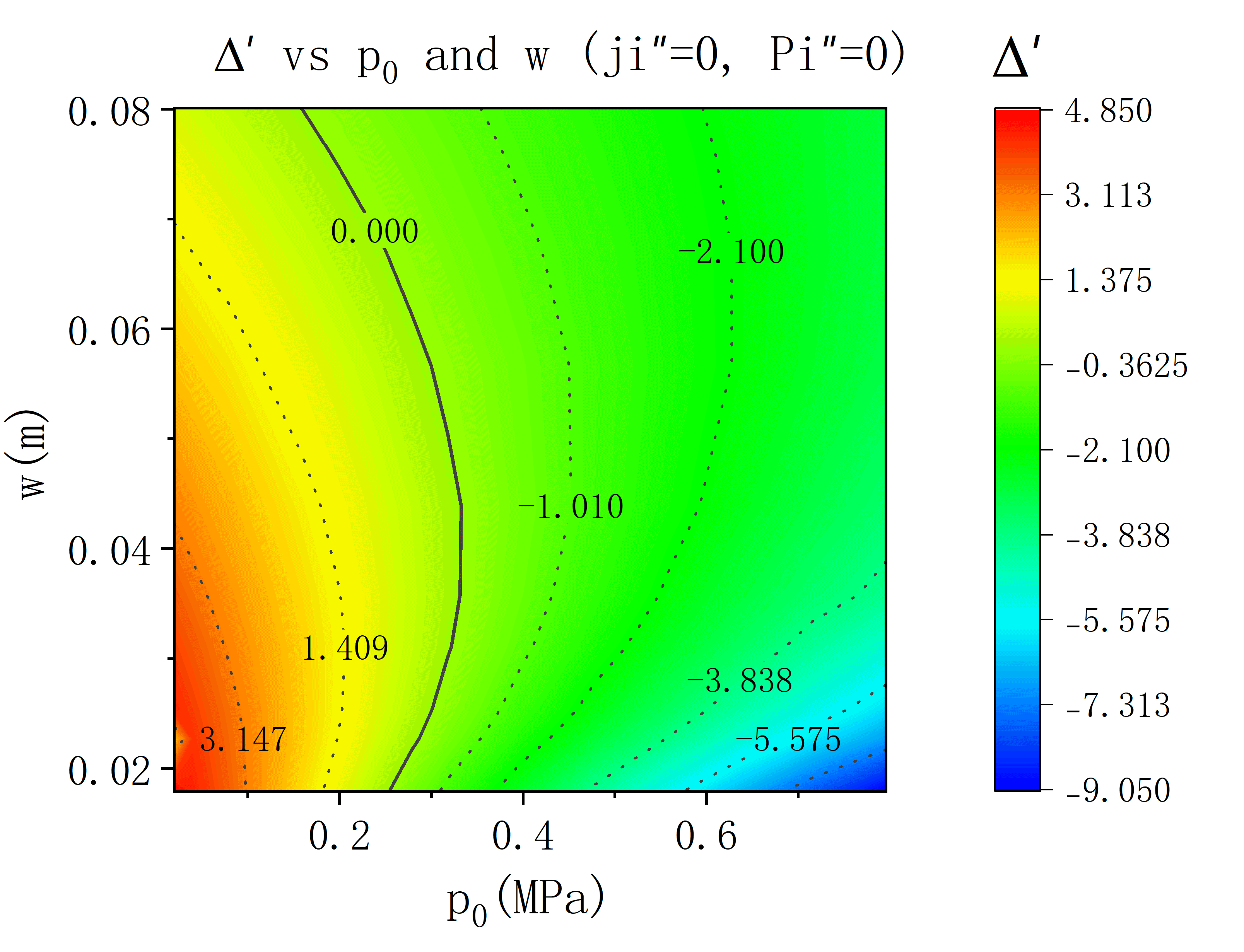}
	\\
	(c)
	\caption{(a) Value of $\gD' $ vary by semi-width and perturbations $j_i''$, with given general pressure $p_0 = 0.239 $MPa, for 2/1 mode and
		edge safety factor $q(a) = 3$ and initial parameter $f_1 = 0.5, f_2 = 0.35$. (b) A line of $j_i'' = 0$ shows that with p(0) the $\gD' $ is not monotonous.
		(c)$\gD'$ changes with the general pressure and semi-width of the island at no current density or pressure gradient perturbation.}
	\label{fig:profileofjwp}
\end{figure*}

Adding finite pressure results in the stabilization of small islands due to the Glasser effect \cite{glassereffect} as is shown in Fig.\,\rfq{fig:profileofjwp}, where a simple flattened pressure profile within the island is considered and we use $m/n=2/1$, island semi-wide $w=0.0801m$, the initial parameter $f_1 = 0.5, f_2 = 0.35$ to set the mode. Comparing Fig.\,\ref{fig:profileofjwp}(a) and  Fig.\ref{fig:profileofjnop}(b), it is apparent that a strong stabilizing effect exists for the small island limit, which is caused by the mode structure modification due to finite pressure gradient near the resonant surface in toroidal geometry \cite{hu_zakharov_2015}. Such pressure contribution vanishes in the large island limit, as is shown in Fig.\,\ref{fig:profileofjwp}(a) \& (b), resulting in a linearly stable but non-linearly unstable scenario. As the pressure increases the island becomes more and more stable, even with a finite island there comes a threshold in $p(0)$ beyond which there is no unstable island, as is shown in Fig.\,\ref{fig:profileofjwp}(c). Note that here the current perturbation from the neo-classical effect is not included, otherwise the island would still be unstable from the decrease of the bootstrap current.

\subsection{Additional contribution from the pressure profile modification}

\begin{figure*}
	\centering
	\noindent
	\btbl{cc}
	\parbox{3in}{
		\includegraphics[scale=0.4]{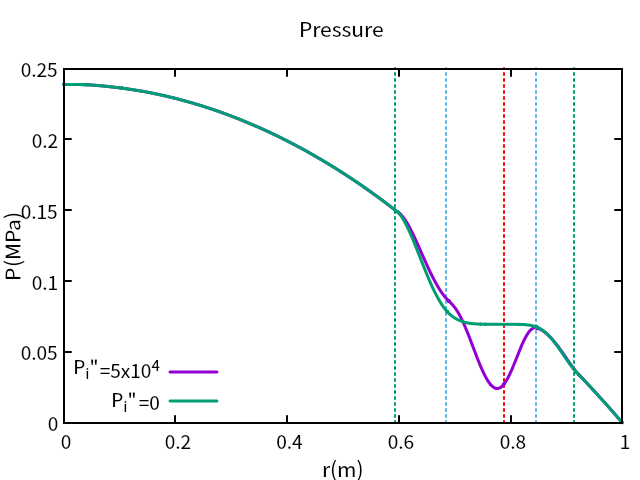}
	}
	&\quad
	\parbox{3in}{
		\includegraphics[scale=0.4]{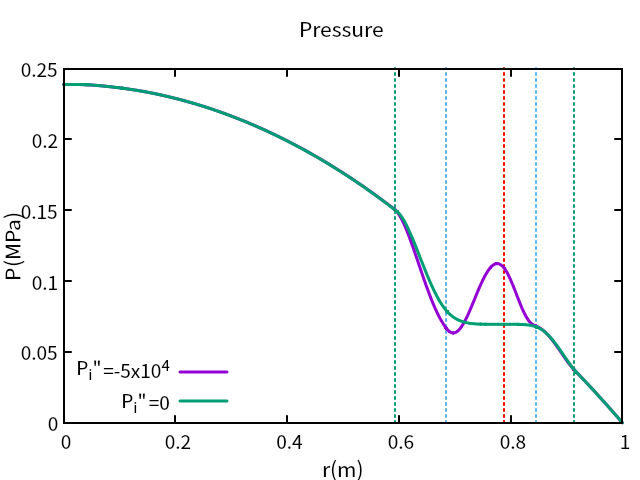}
	}
	\\
	(a)&(b)
	\etbl
	\caption{The pressure of the $2/1$ mode with setting island semi-width to 0.0801m and initial parameter $f_1 = 0.5, f_2 = 0.35$. The five vertical lines still indicate the five zones of the island, and the green line in each figure is the pressure gradient for perturbation is zero and the purple line is for the profile calculated with (a) perturbation $P_i''$ to 50000; (b) set the perturbation $P_i''$ to -50000. }
	\label{fig:profileofp1}
\end{figure*}

\begin{figure*}
	\centering
	\noindent
	\btbl{cc}
	\parbox{3in}{
		\includegraphics[scale=0.3]{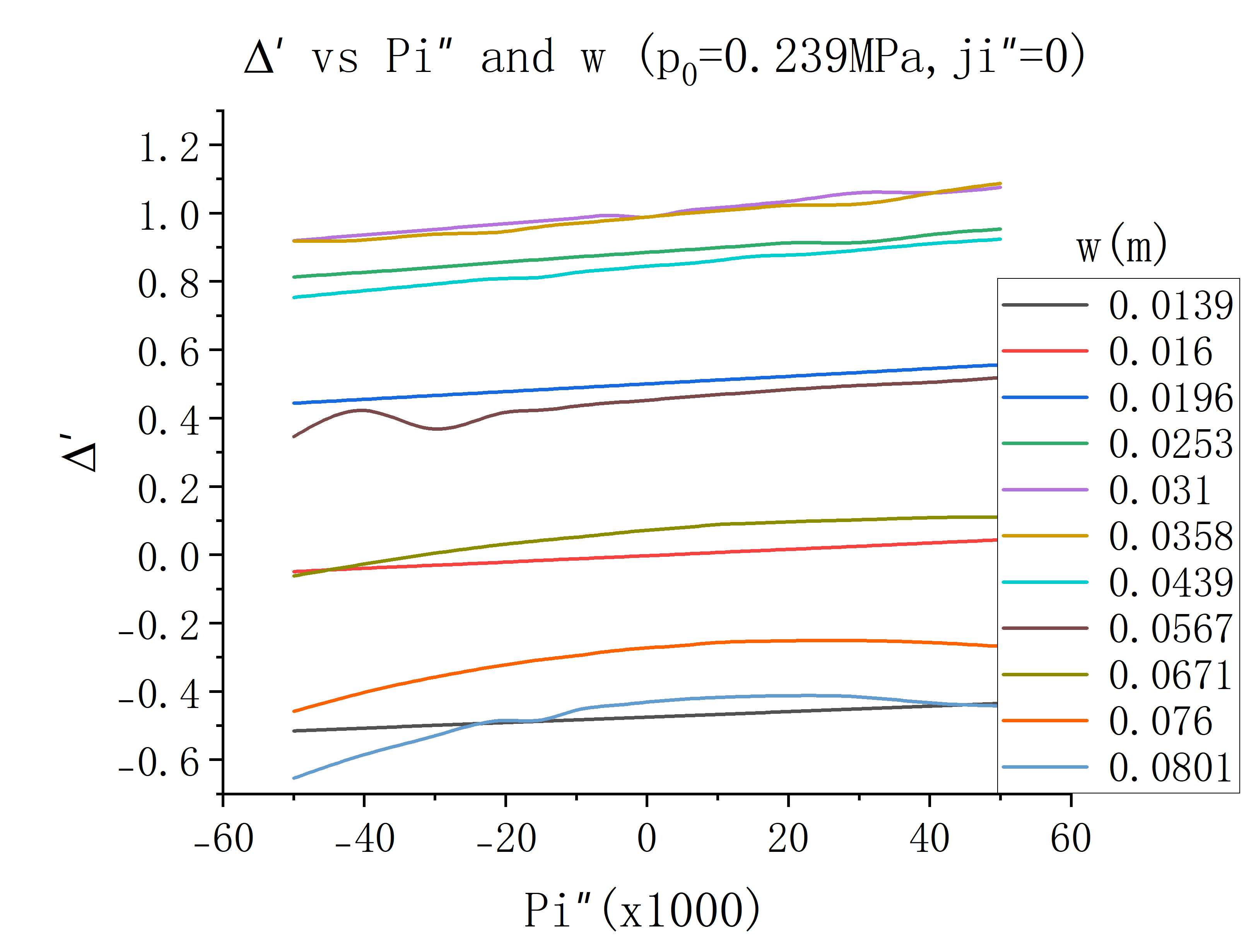}
	}
	&\quad
	\parbox{3in}{
		\includegraphics[scale=0.3]{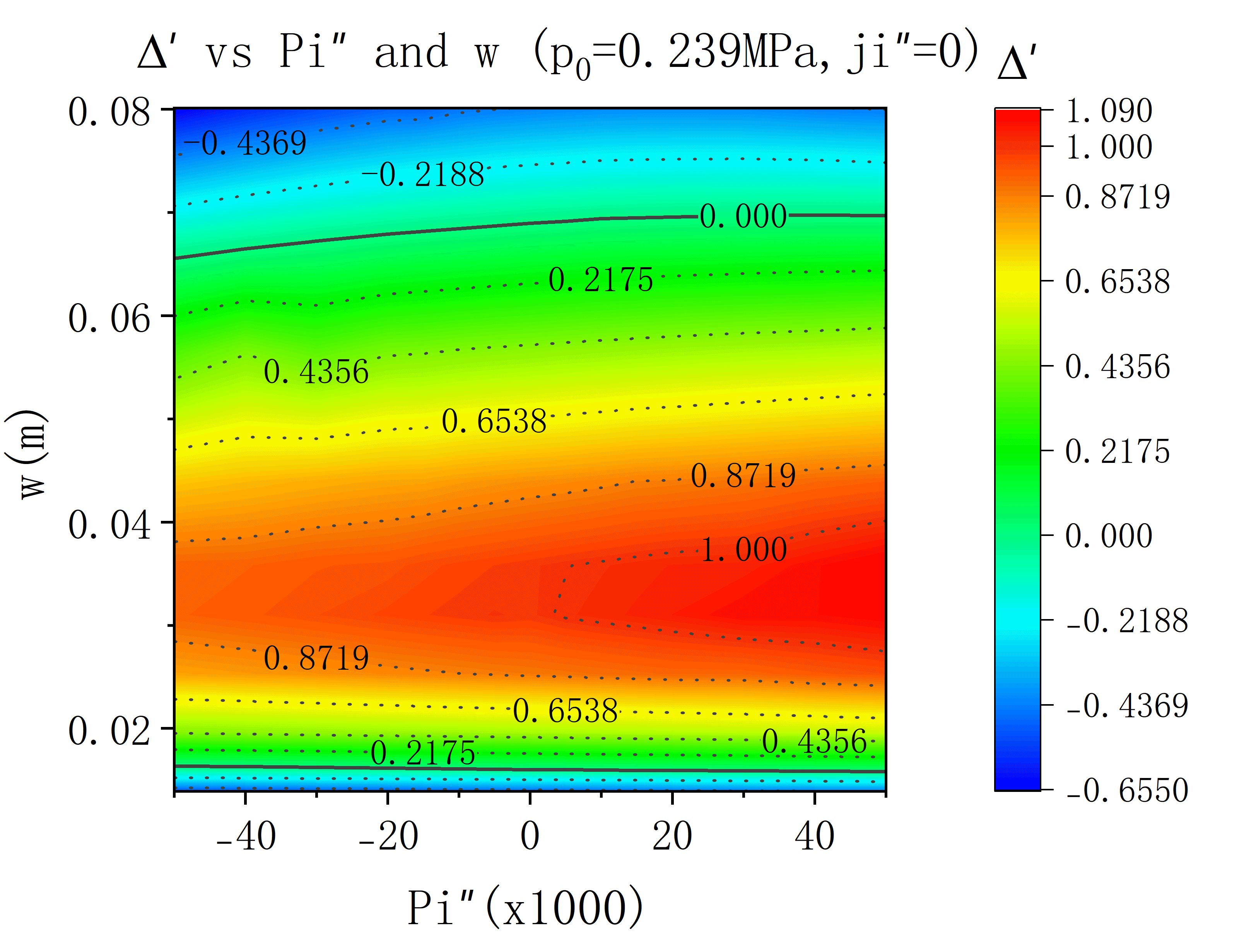}
	}
	\\
	(a)&(b)
	\\
	\parbox{3in}{
		\includegraphics[scale=0.3]{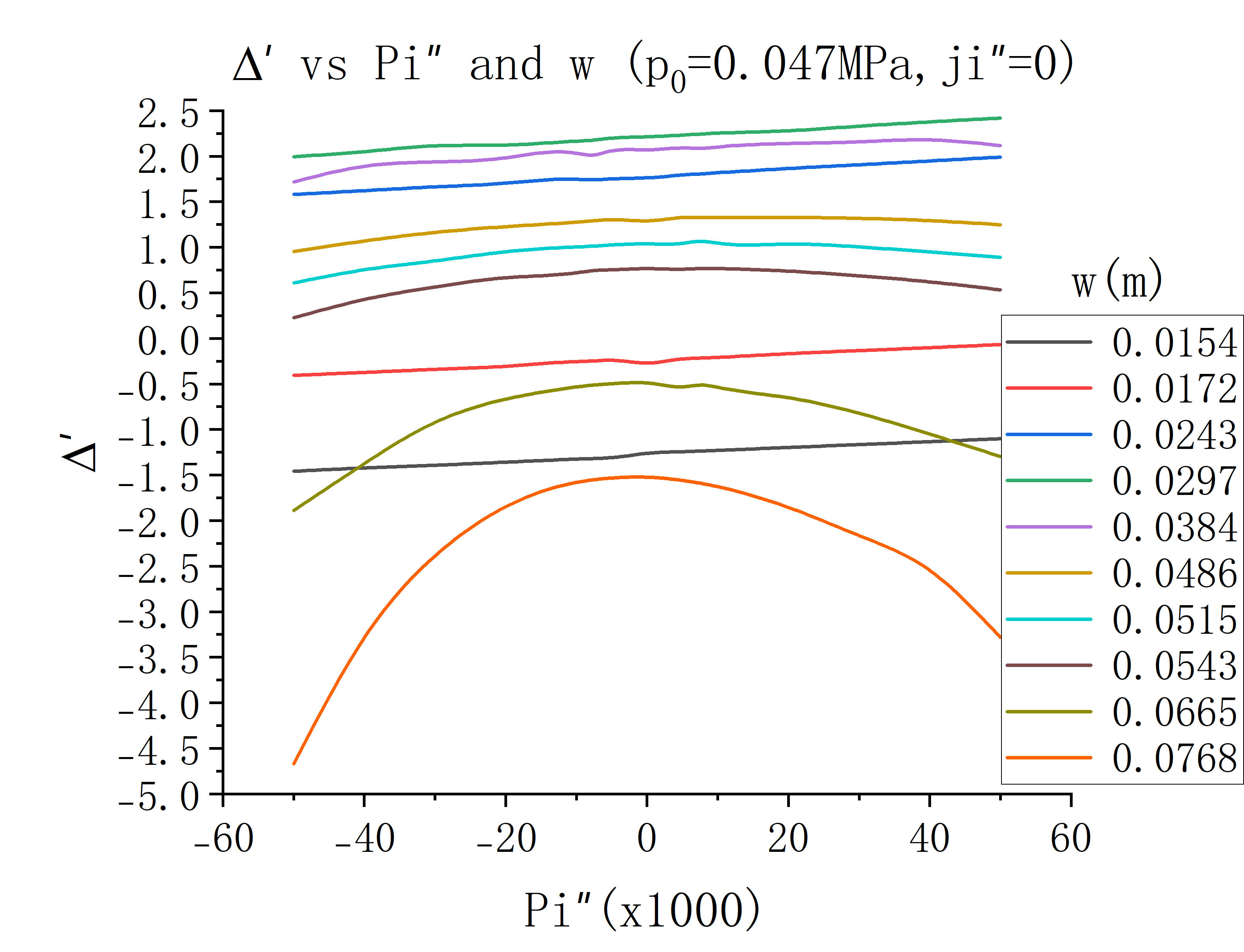}
	}
	&\quad
	\parbox{3in}{
		\includegraphics[scale=0.3]{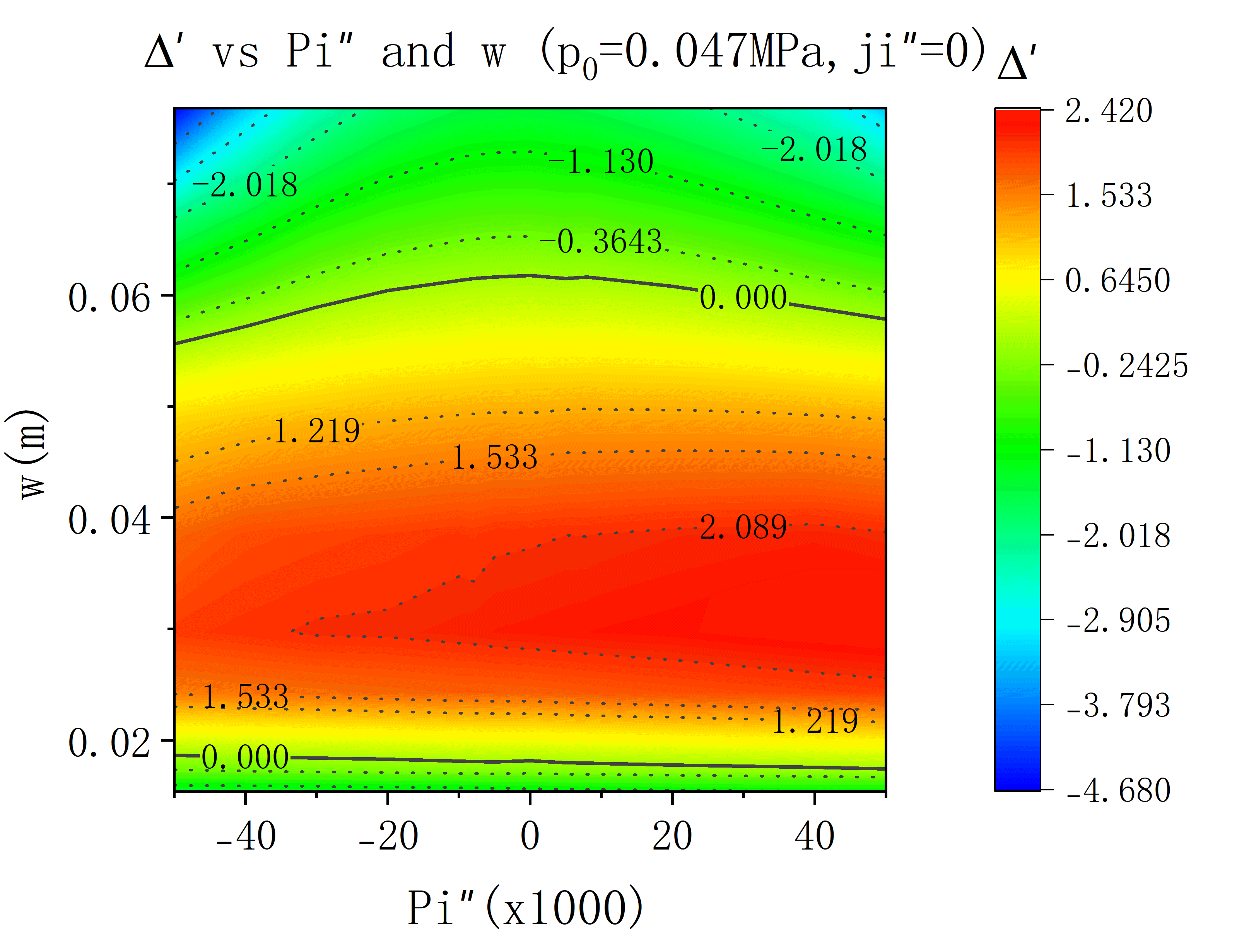}
	}
	\\
	(c)&(d)
	\\
	\parbox{3in}{
		\includegraphics[scale=0.3]{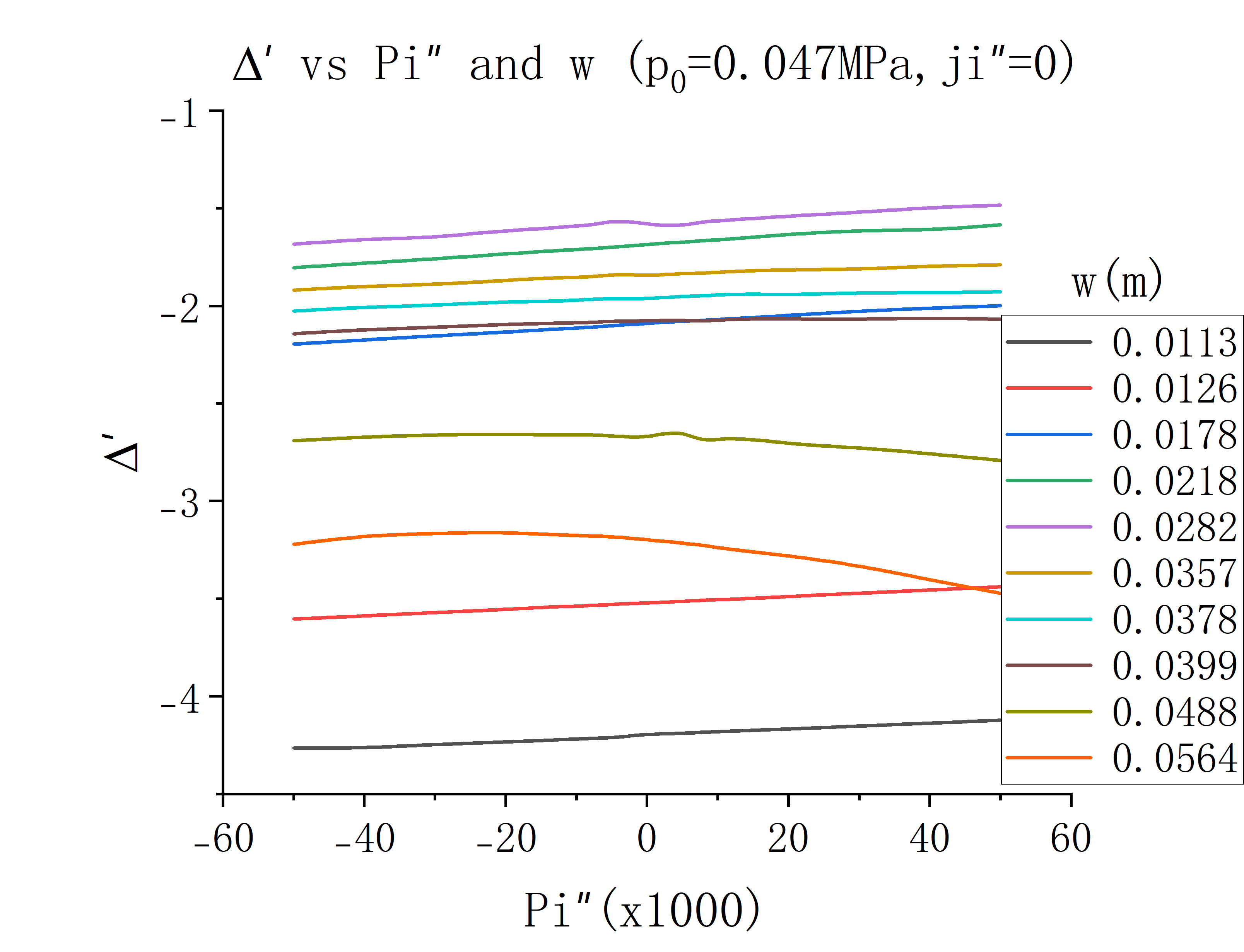}
	}
	&\quad
	\parbox{3in}{
		\includegraphics[scale=0.3]{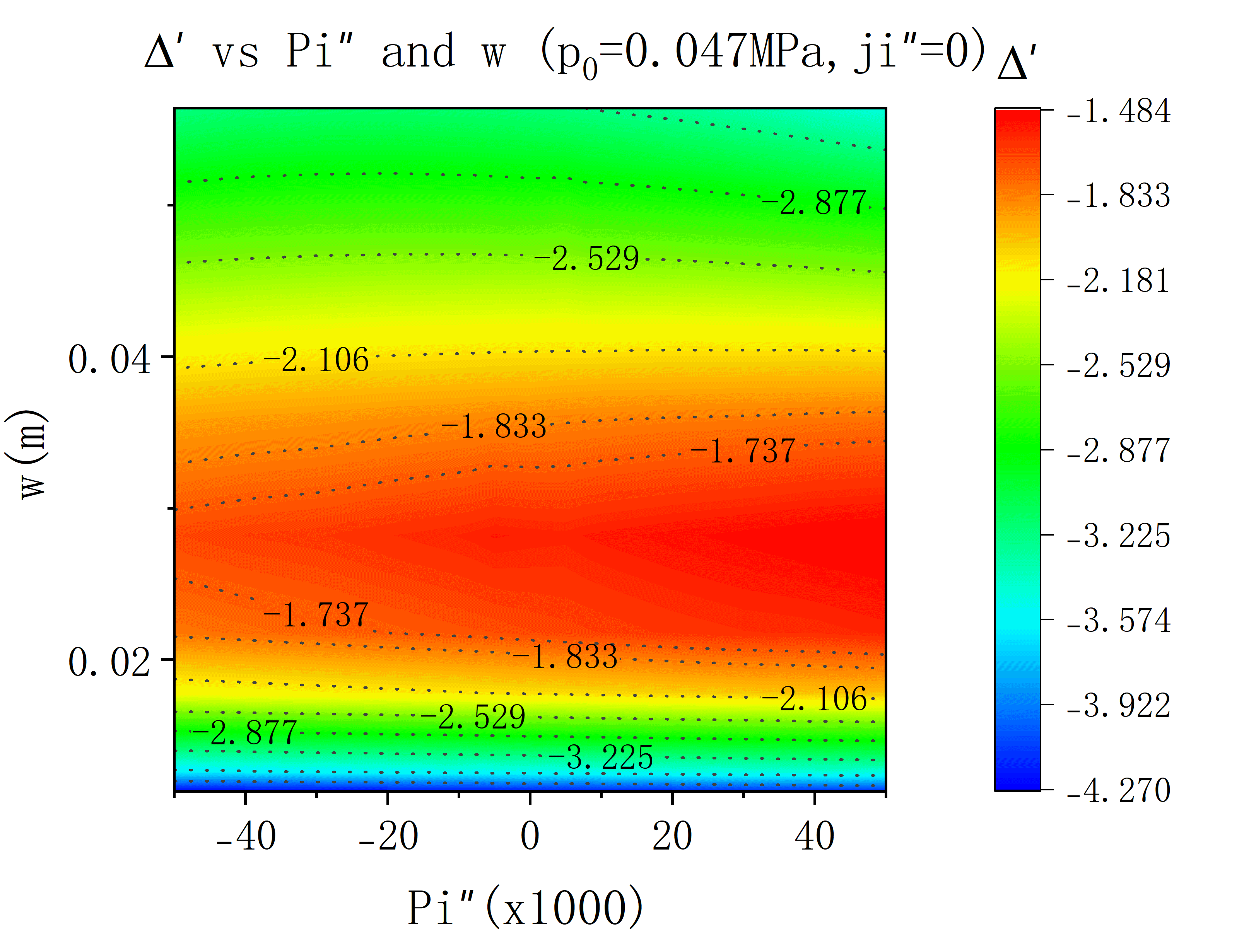}
	}
	\\
	(e)&(f)
	\etbl
	\caption{Value of $\gD' $ vary by semi-width $w$ and perturbations $P_i''$. (a) and (b) are results of 2/1 mode with edge safety factor $q(a) = 3$ and initial parameter $f_1 = 0.5, f_2 = 0.35$; 
		(c) and (d) are results of 2/1 mode with edge safety factor $q(a) = 3$ and initial parameter $f_1 = -0.30, f_2 = -0.20$; (e) and (f) are results of 3/1 mode with edge safety factor $q(a) = 3.95$ and initial parameter $f_1 = -0.30, f_2 = -0.20$.}
	\label{fig:profileofp2}
\end{figure*}

We now consider the stability impact from pressure bumps or holes within the island. To this purpose, we consider a simple flattened current density profile within the island but finite $P_i''$ exists. An example of such pressure modification is shown in Fig.\ref{fig:profileofp1} where Fig.\ref{fig:profileofp1}(a) and (b) are the pressure hole and bump case respectively for the $2/1$ mode case. The stability criterion as a function for various island semi-width $w$ and $P''_i$ is shown in Fig.\ref{fig:profileofp2} for both the $2/1$ mode case with $q(a) = 3$ and the $3/1$ mode case with $q(a) = 4.5$. Here, Fig.\ref{fig:profileofp2}(a) \& (b) show the scan for $2/1$ mode with $f_1=0.5$, $f_2=0.35$, Fig.\ref{fig:profileofp2}(c) \& (d) show that for $2/1$ mode with $f_1=-0.3$, $f_2=-0.2$ and Fig.\ref{fig:profileofp2}(e) \& (f) show that for $3/1$ mode with $f_1=-0.3$, $f_2=-0.2$.
In Fig.\ref{fig:profileofp2}(a), (c) \& (e), the pressure modification effect for a series of fixed island semi-width are shown. In the small island limit, the pressure bump (negative $P''_i$) is found to stabilize the island while the pressure hole (positive $P''_i$) is found to destabilize the island, although their impact is weak due to the cancellation between the odd parity contribution of the $\left(r^2-r_s^2\right)P$ term discussed in Section \ref{s:Qusitheory}. In the large island limit, the pressure impact becomes non-monotonic. Such complicated behaviour could be because of the mode structure, thus its gradient and consequently the island asymmetry, experiencing nonlinear interplay with the pressure profile modification, as the latter depends on the former while also affecting the former. Further, the scan of the island size and the pressure modification for the same cases are shown in Fig.\ref{fig:profileofp2}(b), (d) \& (f). In all cases, a ``ridge-like'' pattern can be seen going from small island to large island, so that the finite sized island could be more unstable compared with linear islands until they tend to saturate. It can also be seen that the $\gD'$ contour at the small island limit is almost unaffected by the change of $P''_i$ while at the large island limits the effect is more obvious, confirming our previous conjecture that the island asymmetry at the large island limit tends to produce stronger pressure gradient effect. Fig.\ref{fig:profileofp2}(b), (d) \& (f) also confirm that in the large island limit, the pressure modification within the island region no longer shows a monotonic impact on stability. Instead, too large pressure bumps or holes could both result in a more stable island.

\begin{figure*}
	\centering
	\noindent
	\btbl{cc}
	\parbox{3in}{
		\includegraphics[scale=0.3]{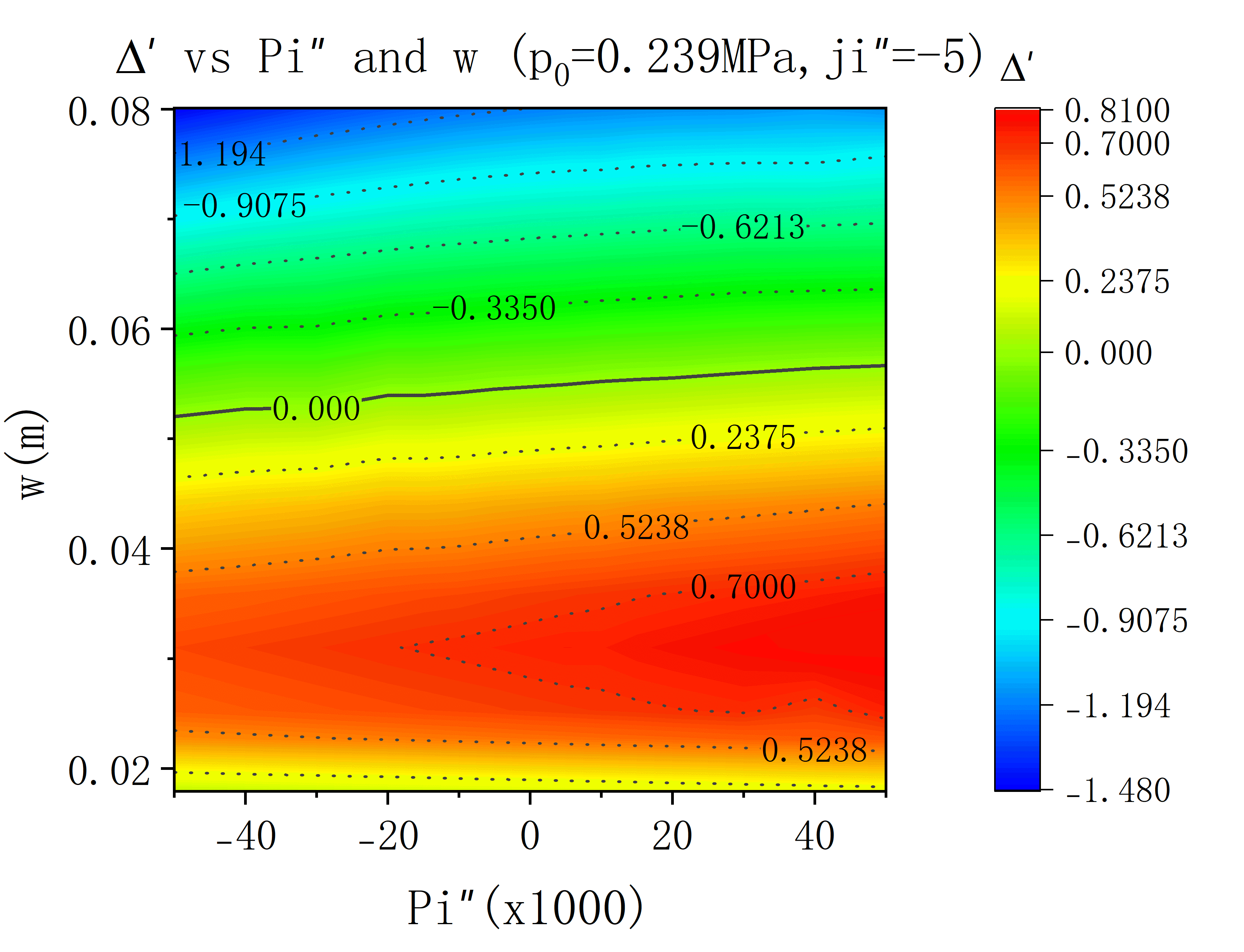}
	}
	&\quad
	\parbox{3in}{
		\includegraphics[scale=0.3]{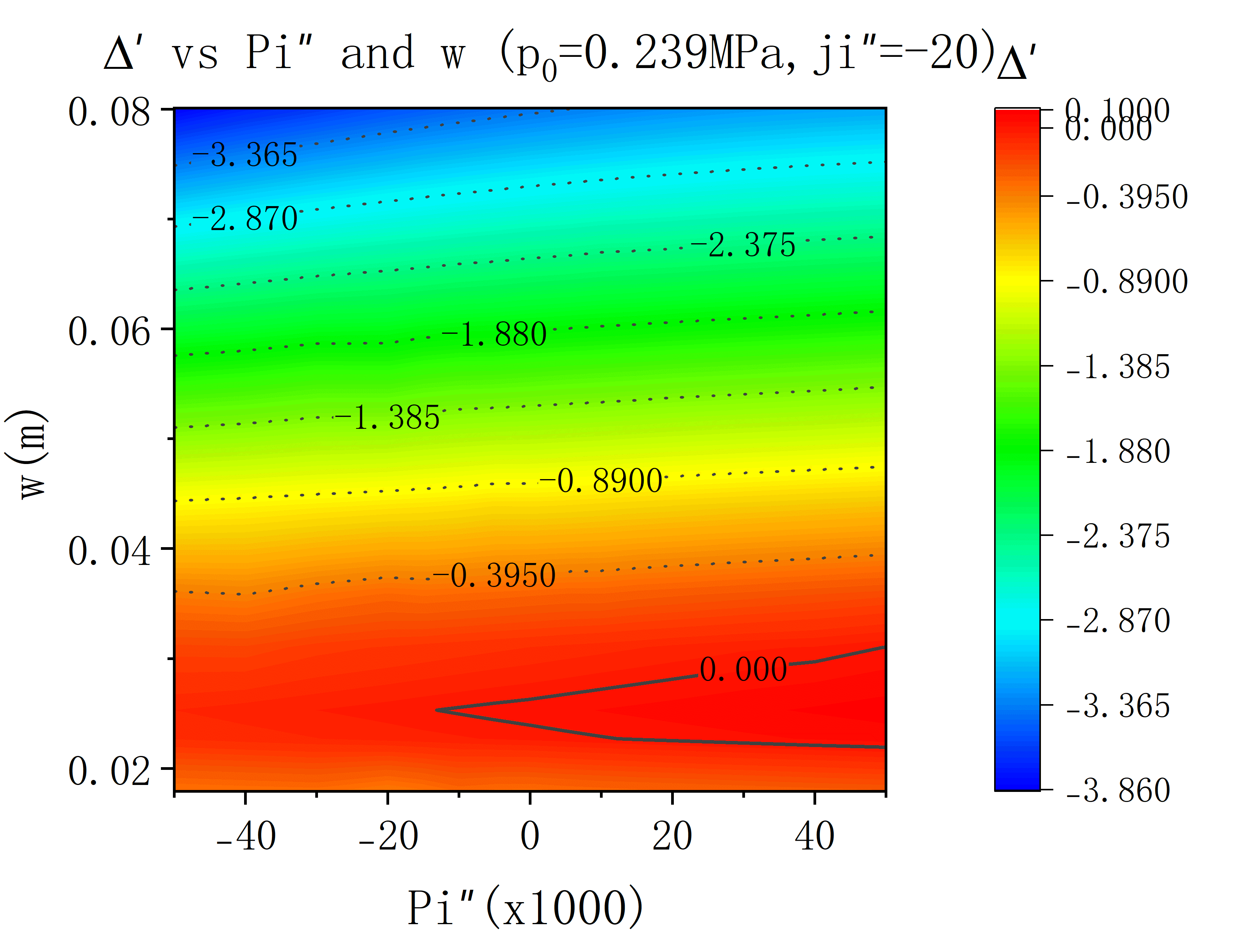}
	}
	\\
	(a)&(b)
	\\
	\parbox{3in}{
		\includegraphics[scale=0.3]{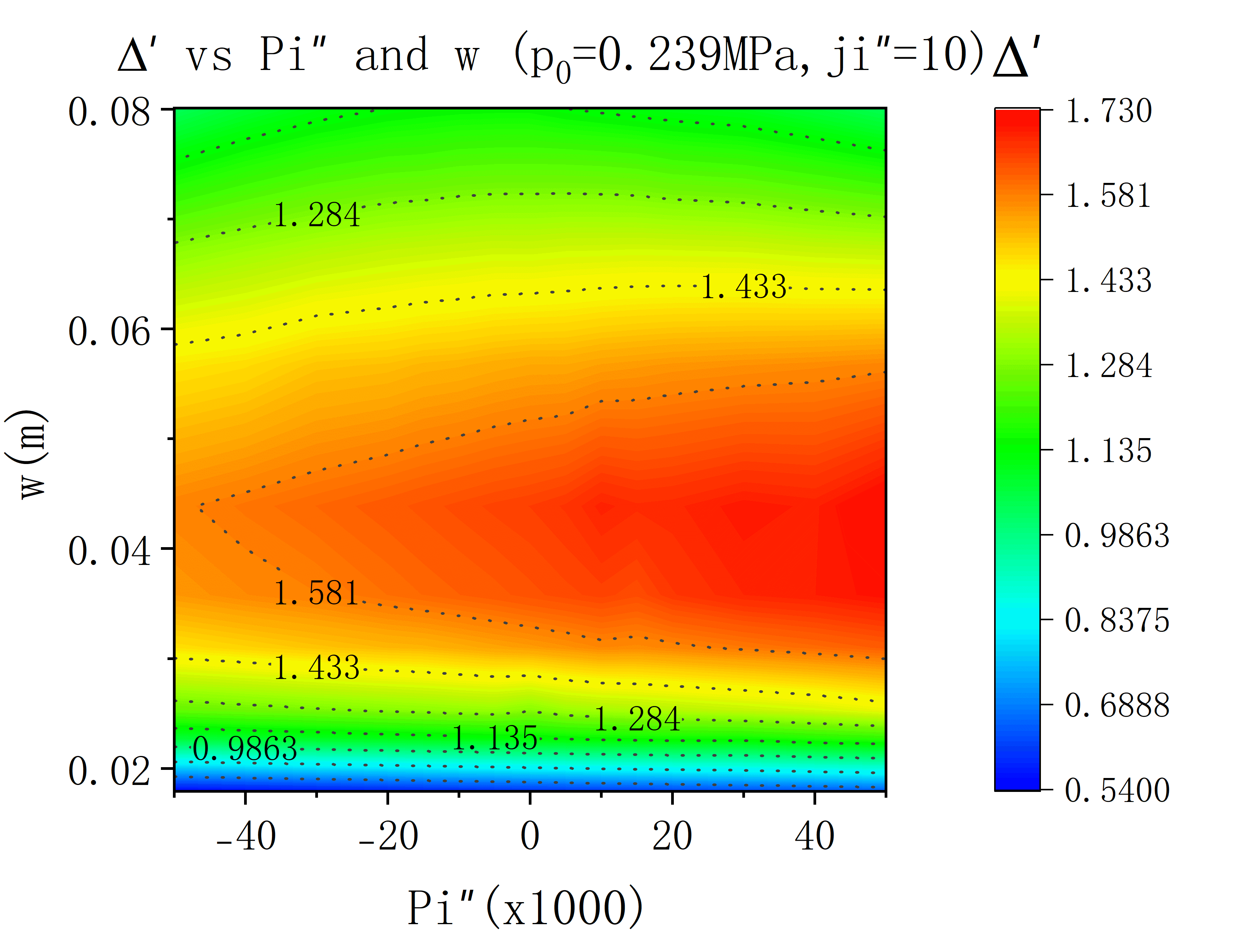}
	}
	&\quad
	\parbox{3in}{
		\includegraphics[scale=0.3]{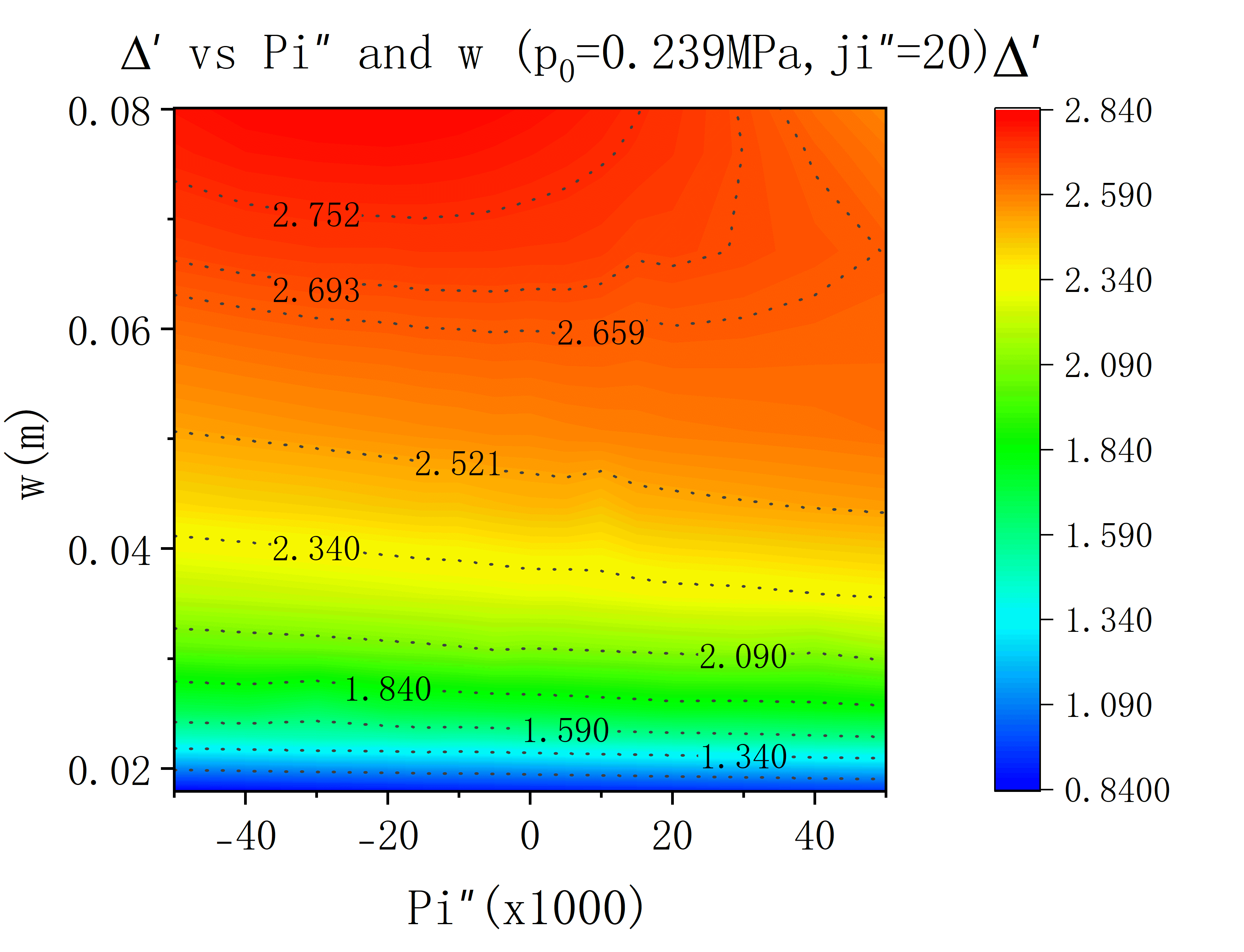}
	}
	\\
	(c)&(d)
	\etbl
	\caption{Change of $\gD' $ figure at given current density perturbation $j_i''$ in the meantime for 2/1 mode and edge safety factor $q(a)=3$, initial parameter $f_1 = 0.5, f_2 = 0.35$, for (a) the $j_i'' = -5$, (b) the $j_i'' = -20$, (c) the $j_i'' = +10$ and (d) the $j_i'' = +20$.}
	\label{fig:profileofp3}
\end{figure*}

We now proceed to combine the current and the pressure modifications within the island. We consider the island stability with several selected $j_i''$ and calculate $\gD'$ for various island semi-width $w$ and pressure modification $P_i''$. The results of 2/1 mode were plotted as Fig.\ref{fig:profileofp3}. We wish to remind the readers again that, although $j''_i$ is constant for each plot, the current perturbation would increase in strength as the island grows larger, so that we will see in Fig.\ref{fig:profileofp3} the instability could ``runaway'' for strong negative helical current perturbation (positive $j''_i$) as the island grows. In Fig.\ref{fig:profileofp3}(a) \& (b), different values of negative $j''_i$ are chosen, corresponding to positive helical current (negative $j''_i$) within the island. In those stabilized cases, the pressure bump (negative $P''_i$) seems to further stabilize the island while the pressure hole does the opposite. For the negative helical current case (positive $j''_i$) shown in Fig.\ref{fig:profileofp3}(c) \& (d), the pressure bump and hole still show a similar impact as is in Fig.\ref{fig:profileofp3}(a) \& (b) for small island width. For the large island width, however, their influence is more complicated and non-monotonic, as both strong pressure bumps and holes could tend to stabilize the island. Such behavior is especially apparent in the large $P''_i$, large island regime in Fig.\ref{fig:profileofp3}(d).
We wish to emphasize again that the bootstrap current contribution should be included within the current modification $j''_i$ in the above analysis, so that any pressure contribution discussed here is in addition to its impact on the current profile.

\begin{figure*}
	\centering
	\noindent
	\btbl{cc}
	\parbox{3in}{
		\includegraphics[scale=0.3]{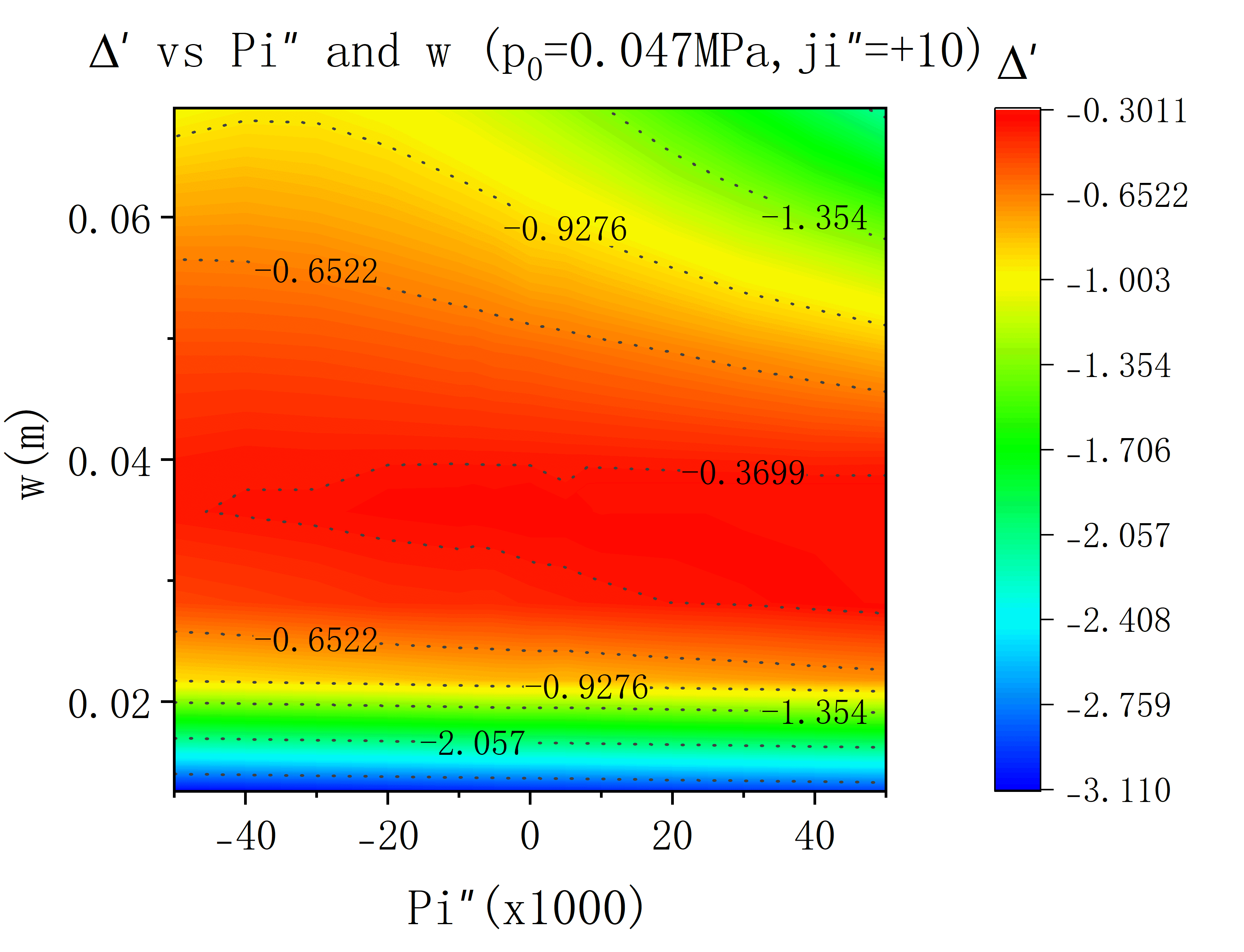}
	}
	&\quad
	\parbox{3in}{
		\includegraphics[scale=0.3]{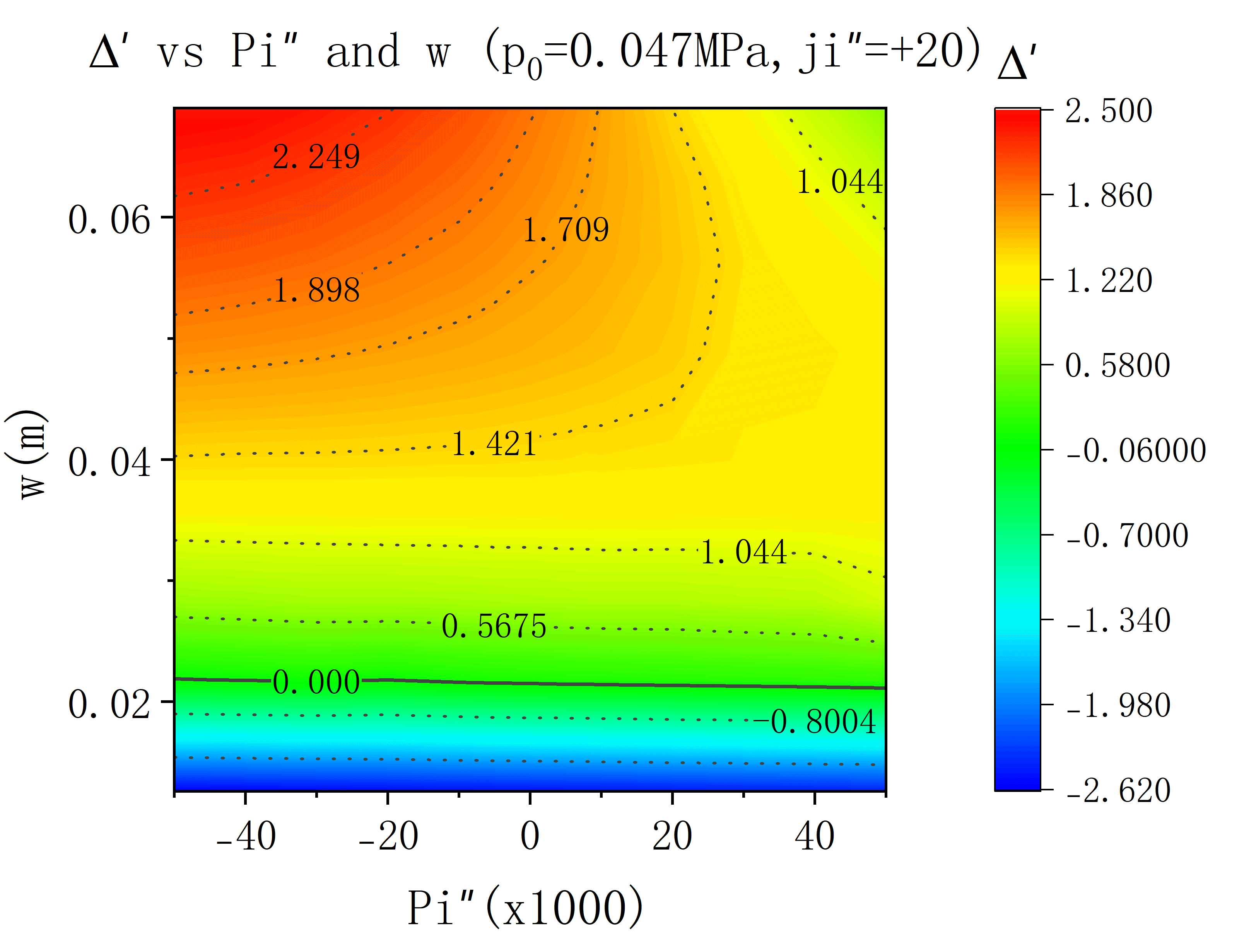}
	}
	\\
	(a)&(b)
	\etbl
	\caption{Change of $\gD' $ figure at given current density perturbation $ji''$ in the meantime for 3/1 mode and edge safety factor $q(a)=4.5$, initial parameter $f_1 = -0.30, f_2 = -0.20$, for (a) the $ji'' = +10$ and (b) the $ji'' = +20$.}
	\label{fig:profileofp4}
\end{figure*}

A similar scan of the $3/1$ mode is plotted in Fig.\ref{fig:profileofp4}. Since the 3/1 mode is stable enough, we don't consider applying negative $j_i''$,
which only decrease $\gD'$. With small island width, the $3/1$ mode shares the same trend of destabilizing the island with increasing $P''_i$ while the same non-monotonic feature appears in the large island limit, as can be seen by comparing Fig.\ref{fig:profileofp4}(a) \& (b) with Fig.\ref{fig:profileofp3}(c) \& (d). Similar ``ridge-like'' feature in $\gD'$ exists for the small current hole case ($j''_i=10$), as can be seen by comparing Fig.\ref{fig:profileofp4}(a) and Fig.\ref{fig:profileofp3}(c). With a stronger current hole ($j''_i=20$), the large $P''_i$ is found to stabilize the island in the large island limit for the $3/1$ mode, consistent with the $2/1$ mode trend. We thus anticipate this to be a common feature that, for a large island, a significant pressure hole would tend to mitigate the destabilizing effect of a current hole within the island, while it contributes to instability in the small island cases.

\begin{figure*}
	\centering
	\noindent
	\btbl{cc}
	\parbox{3in}{
		\includegraphics[scale=0.3]{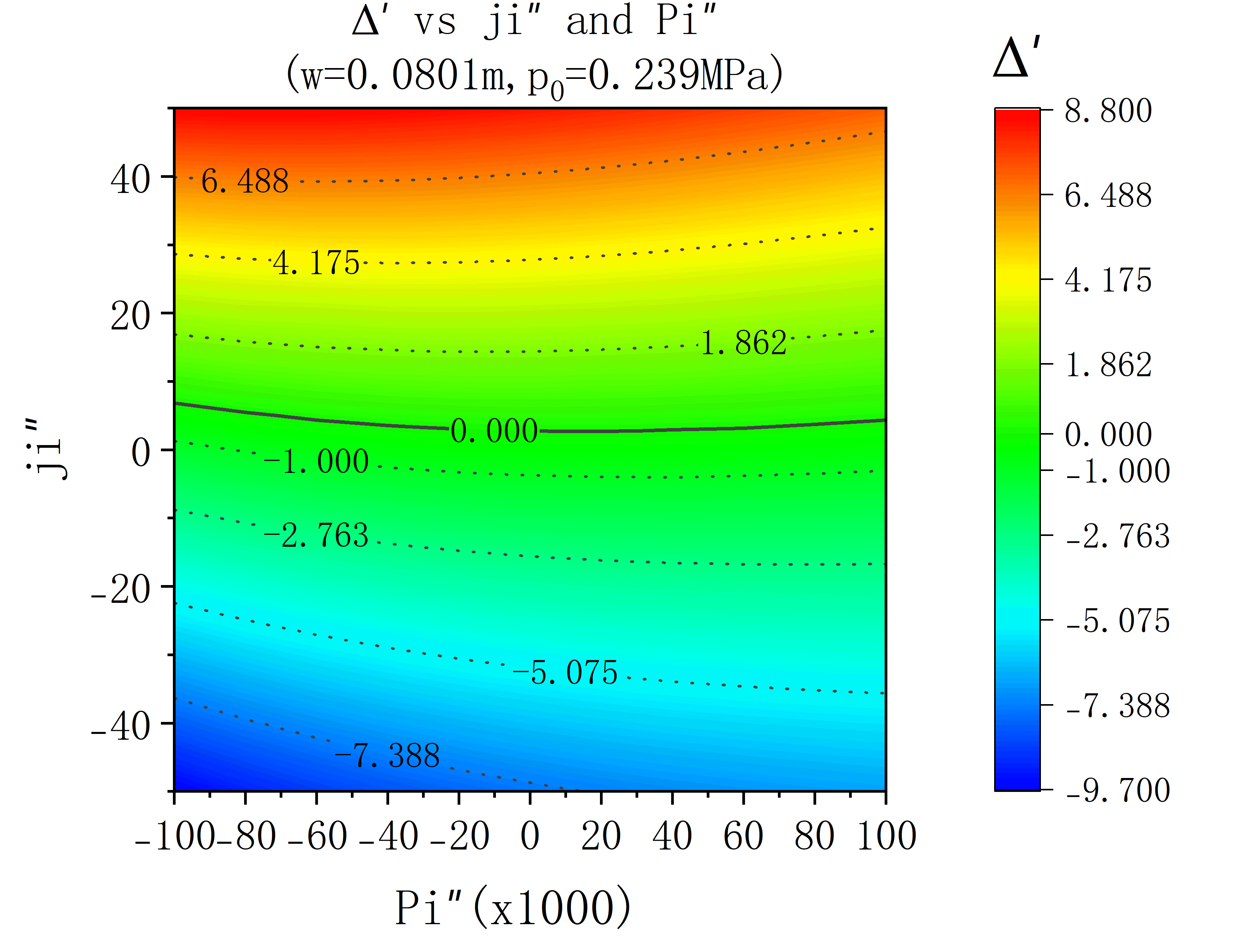}
	}
	&\quad
	\parbox{3in}{
		\includegraphics[scale=0.3]{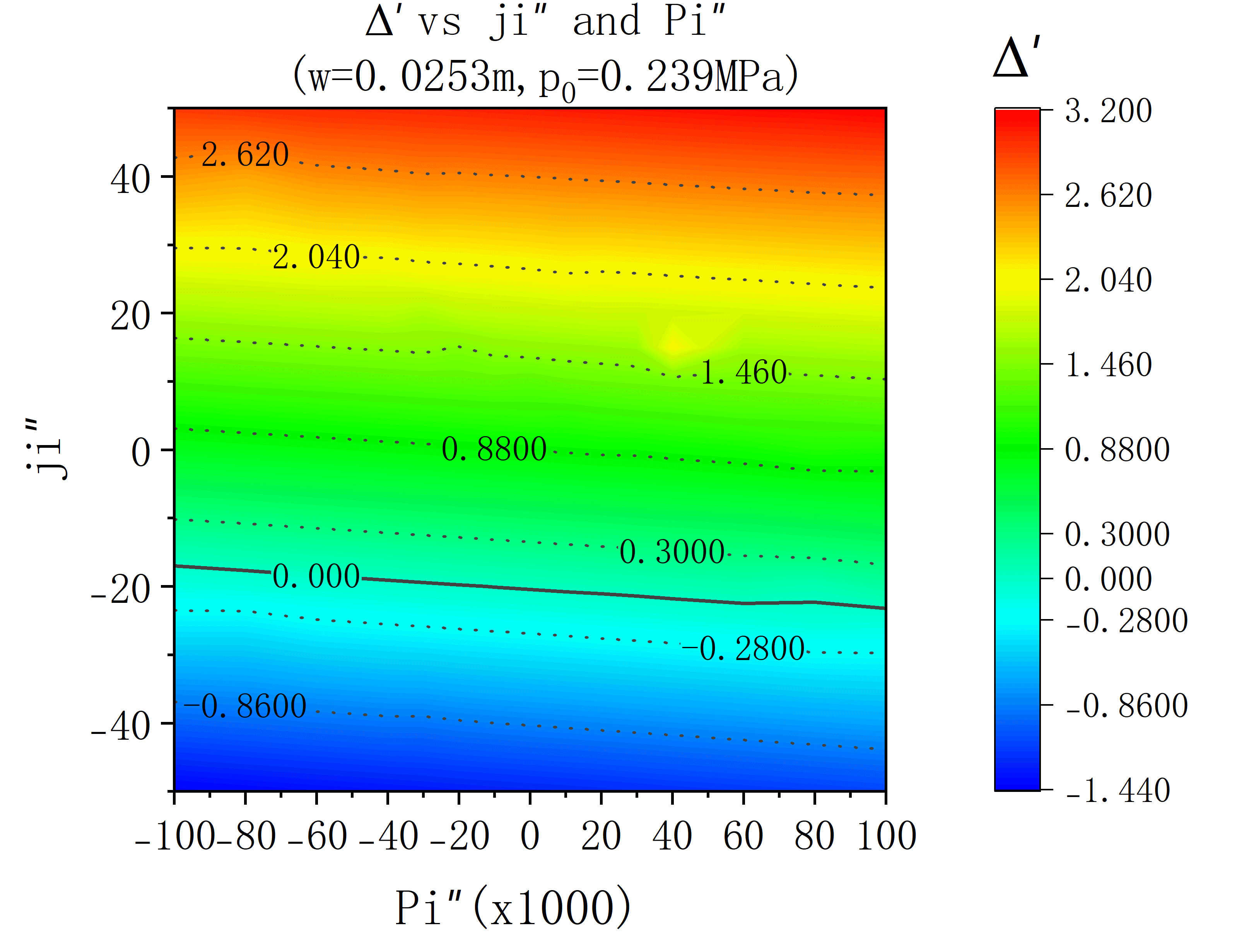}
	}
	\\
	(a)&(b)
	\etbl
	\caption{Change of $\gD' $ figure at given semi-width in the meantime for 2/1 mode and edge safety factor $q(a)=3$, initial parameter $f_1 = 0.5, f_2 = 0.35$, for (a) the semi-width $w = 0.0801m$ and (b) the semi-width $w = 0.0253m$.}
	\label{fig:profileofjipi}
\end{figure*}

The above mentioned trend could be more directly shown by considering a scan of different $j''_i$ and $P''_i$ with constant island width. We choose two different island semi-width $w=0.0235m$ and $w=0.0801m$, the results are plotted in Fig.\ref{fig:profileofjipi}. For the large island case shown in Fig.\ref{fig:profileofjipi}(a), the $P''_i$ stability impact is once again seen as non-monotonic, as large positive $P''_i$ is seen to mitigate the destabilizing effect of a helical current hole (positive $j''_i$), and large negative $P''_i$ is also found to stabilize the island. For the smaller island case shown in Fig.\ref{fig:profileofjipi}(b), the $P''_i$ contribution to island stability is mostly monotonic, with pressure bump consistently stabilizing the island and pressure hole consistently destabilizing the island. These different behaviours in different island width regimes shown in Fig.\ref{fig:profileofjipi} are consistent with what we discussed above regarding the pressure profile modification stability influence.

\section{Conclusion and discussion}
\label{s:Conclusion}

The stability analysis of finite size island with current and pressure profile modifications near the O-point is carried out in this study using a quasi-linear perturbed equilibrium approach. The helical Grad-Shafranov equation is considered and the mode structure of the principle harmonic of the perturbed helical flux is solved up to the resonant surface to obtain the island stability criterion $\gD'$, where the PS current term from the curvature and pressure is neglected, because we use a small aspect-ratio. The profile modification within the island is introduced by considering the 2D profiles and extracting their zeroth and principle harmonic to feed into the RHS of the helical Grad-Shafranov equation. Among the RHS terms, the helical current perturbation contributes consistently to the island's stability, with positive helical current perturbation always stabilizing the island while the negative one does the opposite. These findings are in agreement with the previous constant $\gy$ analysis. The pressure contribution is more complicated due to the cancellation of its odd parity contribution in the second derivative of the perturbed helical flux. In the small island regime, the pressure modification contribution is monotonic, and the pressure bump is found to stabilize the island while the pressure hole does the opposite, but their overall stability impact is weak due to the aforementioned cancellation. In the large island regime, the pressure impact becomes non-monotonic, especially when we combine the pressure and current modifications. Both large pressure bumps and holes are found to stabilize the island in this regime.

Based on our conclusions above, for example, one would expect that the local heating by electron cyclotron wave within the island would result in an additional stabilization effect apart from the current drive caused by ECCD. Another example is the injection of deuterium pellets into the island O-point, which results in dilution cooling inside the island region. The cooling itself would result in resistive helical current expulsion from the island O-point, thus destabilizing. But the low temperature island region would also cause heat flux to flow into the island O-point, increasing its thermal density and pressure. Based on our result, one would anticipate the pressure bump caused by such an injection to mitigate the destabilization caused by the current expulsion, thus slowing down the island's growth. These are some examples derived from our results above.

We would like to emphasize again that the current and pressure perturbation considered in this study is arbitrary, so long as they can be expressed as a function of the helical flux. 
Some immediate future development would be pursued to obtain a more self-consistent description of the island's stability. For example, the bootstrap current modification could be automatically specified for a given pressure modification, so that the bootstrap contribution can be taken into account self-consistently\cite{Kikuchi1995}. The same is true for the vortices mode contribution within the island which impacts the thermal transport, resulting in pressure profile modification thus bootstrap current modification\cite{Hornsby2010}.
Implementing consistent description of the aforementioned profile modifications will be part of our future works. Another future development of the current work is to retain the principal harmonic of the nonlinear terms involving the other helicity modes in the RHS of Eq.\,(\rfq{eq:Equilibrium10}) to take into the nonlinear coupling with other islands. The non-circular shaping effect of the plasma cross-section can be like-wise be added. Furthermore, the convective term could be added into the force balance equation to take the finite flow effect into account in the equilibrium solution.

Overall, the study carried out here provides insight into the island's response to various disruption avoidance or mitigation scheme such as current drive and local heating caused by cyclotron waves, or local cooling and current expulsion caused by massive material injections. Understandings on the above dynamics would help to achieve more efficient disruption avoidance and mitigation in future machines.

\vskip1em
\centerline{\bf Acknowledgments}
\vskip1em

  The authors thank L.E. Zakharov for fruitful discussion and permission to carry out code development upon his code infrastructure. This work is supported by the National MCF Energy R\&D Program of China under Grant No. 2019YFE03010001 and the National Natural Science Foundation of China under Grant No. 11905004.

\end{document}